%% file: main.tex
\documentclass{article}
\usepackage[a4paper,left=2cm,top=1.5cm,bottom=1.5cm,right=2cm]{geometry}
\usepackage[utf8]{inputenc}
\pdfoutput=1
\usepackage{color}
\usepackage{xcolor}
\usepackage{hyperref}
\definecolor{linkcolour}{rgb}{0,0.2,0.6}
\hypersetup{colorlinks,breaklinks,urlcolor=linkcolour, linkcolor=linkcolour, citecolor=blue}
\usepackage{graphicx}
\usepackage{caption}
\usepackage{subcaption}
\usepackage{multirow}
\usepackage{placeins}
\usepackage{booktabs}
\usepackage{lineno}
\usepackage{authblk}
\usepackage{lineno}
\usepackage{afterpage}
\usepackage{graphicx}
\usepackage{subcaption}
\newlength{\twosubht}
\newsavebox{\twosubbox}
\usepackage{hyperref}

\makeatletter         
\renewcommand\maketitle{
{
\begin{center}
{\huge \@title \par}
\vspace{1cm}
{\large \textbf{The ECFA Early-Career Researchers (ECR) Panel}}
\\[1cm]
{\large \@date}
\\[1cm]
\begin{minipage}{0.82\textwidth}
\normalsize A 2021 study by the ECFA Early-Career Researchers Panel \href{https://arxiv.org/pdf/2107.05739}{(arXiv:2107.05739)} revealed that 71$\%$ of 334 respondents used open-source software tools in their instrumentation work, yet 70$\%$ reported receiving no training for these tools. In response, the Software and Machine Learning for Instrumentation group was formed in the ECFA Early-Career Researchers Panel to assess the accessibility and quality of training programs in machine learning and software for early-career researchers in experimental and applied physics. This group launched a new survey, reaching 174 participants. This report summarises the survey results in detail, and is intended to serve as a guiding document to improve the training programs that are available to early-career researchers. 
\end{minipage}
\end{center}
\vspace{1.5cm}
\begin{flushleft}
{The ECFA Early-Career Researchers (ECR) Panel: \href{mailto:ecfa-ecr-organisers@cern.ch}{ecfa-ecr-organisers@cern.ch}\\[0.5cm]\@author}
\end{flushleft}}}
\makeatother

\RequirePackage[style=numeric,backend=biber,sorting=none]{biblatex}
\begin{document}

\title{Results of the analysis of survey for young scientists on training quality in HEP instrumentation software and machine learning}
\date{\today}

\input{authors.tex}

\maketitle

\clearpage
\section{Introduction}
\input{01_introduction}
\section{Accessibility and quality of schools}
\input{schools_general}

\section{Machine Learning}
\input{02_ML}

\section{Detector simulation}
\input{03_DetSim}
\clearpage

\section{Data Acquisition and detector control systems}
\input{04_DAQ}

\section{Detector electronics}
\input{05_Electronics}

\section{Summary and conclusions}
\input{Conclusion}

\clearpage

\addcontentsline{toc}{section}{Bibliography}
\printbibliography[title=References]

\end{document}

%% file: authors.tex

\author[*,1]{Cecilia Borca}
\author[*,2]{Javier Jiménez Peña}
\author[*,3]{David Marckx}
\author[*,4]{Małgorzata Niemiec}
\author[*,5]{Elisabetta Spadaro Norella}
\author[*,6]{Marta Urbaniak}

\affil[*]{Editor}

\affil[1]{University of Torino and INFN, Turin; Italy}
\affil[2]{Institut de Física d'Altes Energies, Barcelona; Spain}
\affil[3]{Ghent University, Ghent; Belgium}
\affil[4]{Henryk Niewodniczanski Institute of Nuclear Physics Polish Academy of Sciences, Kraków; Poland}
\affil[5]{University of Genoa and INFN, Genoa; Italy}
\affil[6]{University of Silesia in Katowice, Katowice; Poland}

%% file: 01_introduction.tex
A 2021 study by the ECFA Early-Career Researchers Panel \href{https://arxiv.org/pdf/2107.05739}{(arXiv:2107.05739)} revealed that 71$\%$ of 334 respondents used open-source software tools in their instrumentation work, yet 70$\%$ reported receiving no training for these tools. In response, the Software and Machine Learning for Instrumentation group was formed to assess the accessibility and quality of training programs in machine learning and software for early-career researchers in experimental and applied physics. 
\\

A new survey was published by the Software and Machine Learning for Instrumentation group in 2025 to gain new insights and guide future decision-making around training programs. This survey is the focus of this document and was structured as follows. At the outset, we aimed to gain a general understanding of the respondents; therefore, the first section consisted of broad, general questions designed to characterize the surveyed community. The subsequent section, also general in nature, focused on whether participants were aware of the educational offerings provided by the schools and whether they made use of them. The remaining sections were completed only by participants who expressed interest in specific topics, such as machine learning, detector simulation, DAQ and DCS systems, and detector electronics.
\\

\begin{figure}[h!]
\sbox\twosubbox{%
  \resizebox{\dimexpr.99\textwidth-1em}{!}{%
    \includegraphics[height=3cm]{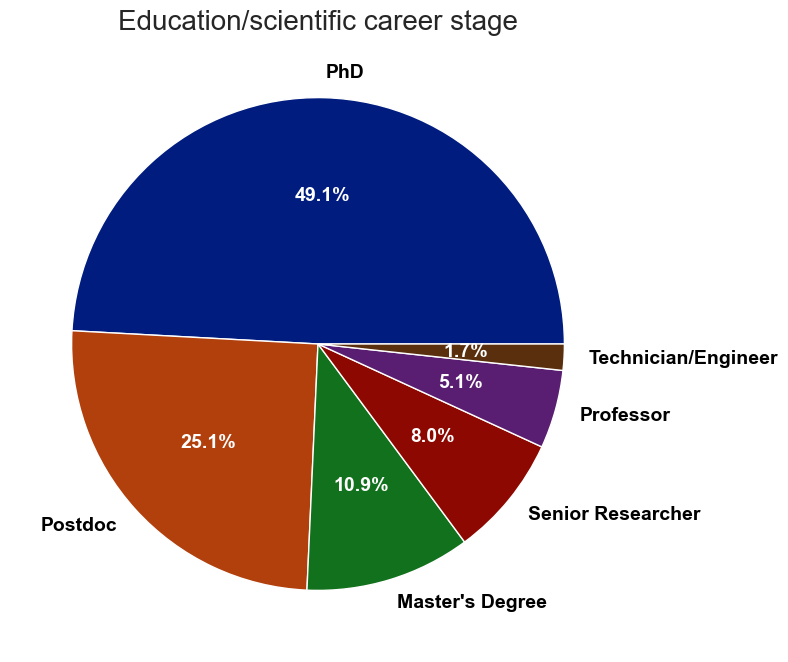}%
    \includegraphics[height=3cm]{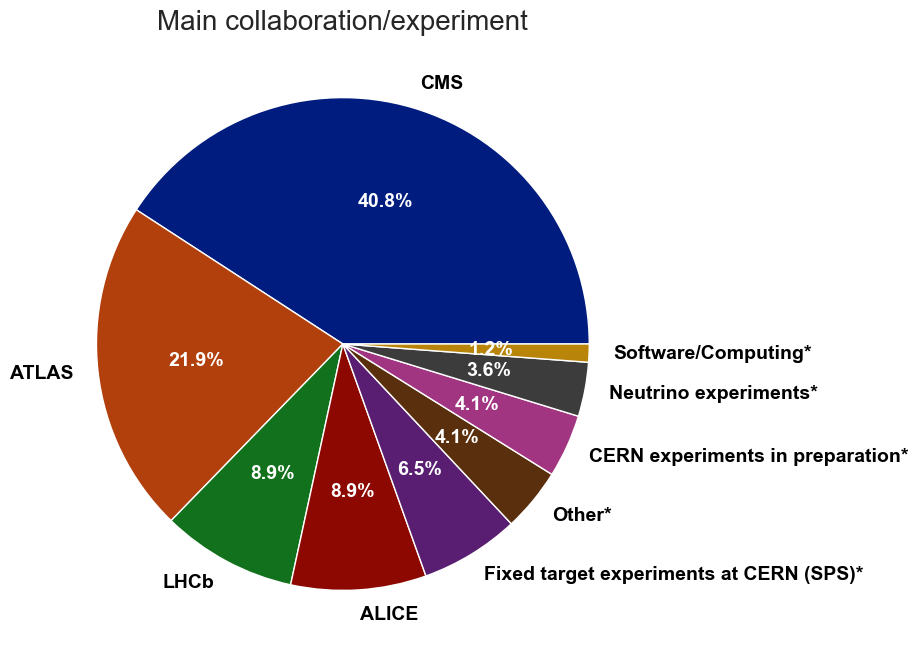}%
  }%
}
\setlength{\twosubht}{\ht\twosubbox}


\centering

\subcaptionbox{\label{fig:in:carStage}}{%
  \includegraphics[height=\twosubht]{figures/introduction/What_is_your_educationscientific_career_stage.png}%
}\quad
\subcaptionbox{\label{fig:in:expe}}{%
  \includegraphics[height=\twosubht]{figures/introduction/Name_of_the_experiment_Insert_only_the_acronym,_if_applicable..png}%
}
\caption{The respondents’ career stage distribution (a). Name of the experiment in which the respondents take part (b).}
\end{figure}

The survey included 174 participants, around 86$\%$ of them work in Europe, the remaining $14\%$ came from numerous other nations across Asia, the Americas, and Africa. Regarding career stage, participants included postdoctoral researchers, master's degree holders, senior researchers, university professors, and technical or engineering staff. The percentage distribution of the respondents’ scientific career stages can be seen in Figure~\ref{fig:in:carStage}. Since the survey was primarily targeted at Early Career Researchers (ECRs), postdoctoral researchers and students make up the majority of the sample. The majority of the surveyed community is involved in large collaborations at the Large Hadron Collider (LHC), each comprising more than 1,000 members, as shown in Figure~\ref{fig:in:expe}. The next largest group, representing 6.5$\%$ of the responses, consists of participants in fixed-target experiments at CERN’s SPS accelerator, such as NA61/SHINE, COMPASS, AMBER, and NA62. Respondents affiliated with research groups or experiments outside CERN constitute 8.9$\%$ of the sample. These primarily include neutrino experiments such as DUNE, T2K, ENUBET, and Hyper-Kamiokande (HK). A smaller category includes participants involved in Software and Computing initiatives (e.g.\ GEANT4, ICSC Spoke~3). Responses labeled as \textit{Other} (single entries) correspond to diverse projects such as Belle II, PADME, CTA, HESS, ePIC, and ALTO/CoMET. Finally, 4.1$\%$ of the respondents reported participation in projects or experiments currently under preparation, including FCC, HI-BEAM, CODEX-b, QUART$\&$T, LUXE, and ILD.  
\\

\begin{figure}[h!]
\sbox\twosubbox{%
  \resizebox{\dimexpr.99\textwidth-1em}{!}{%
    \includegraphics[height=3cm]{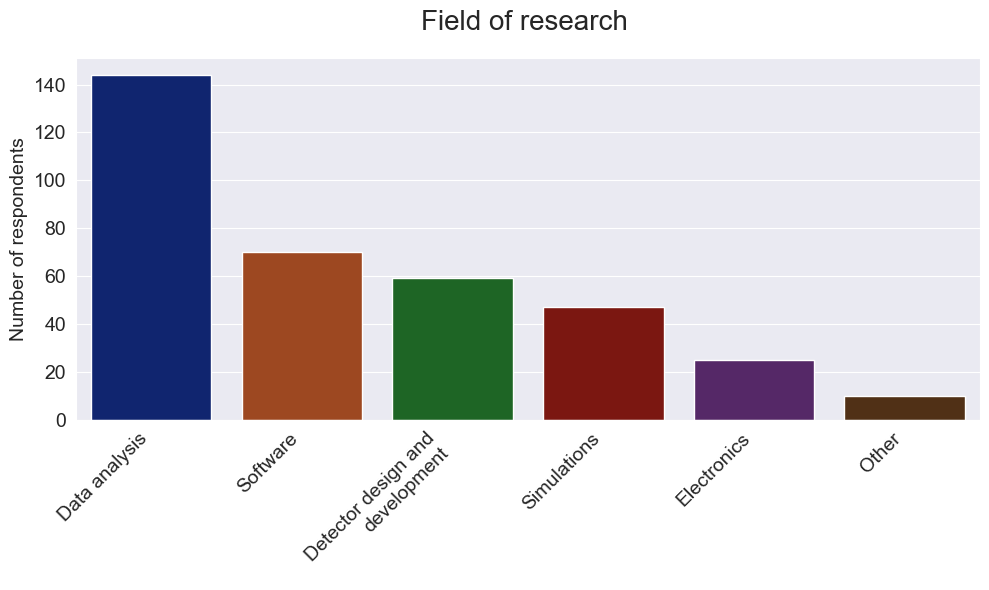}%
    \includegraphics[height=3cm]{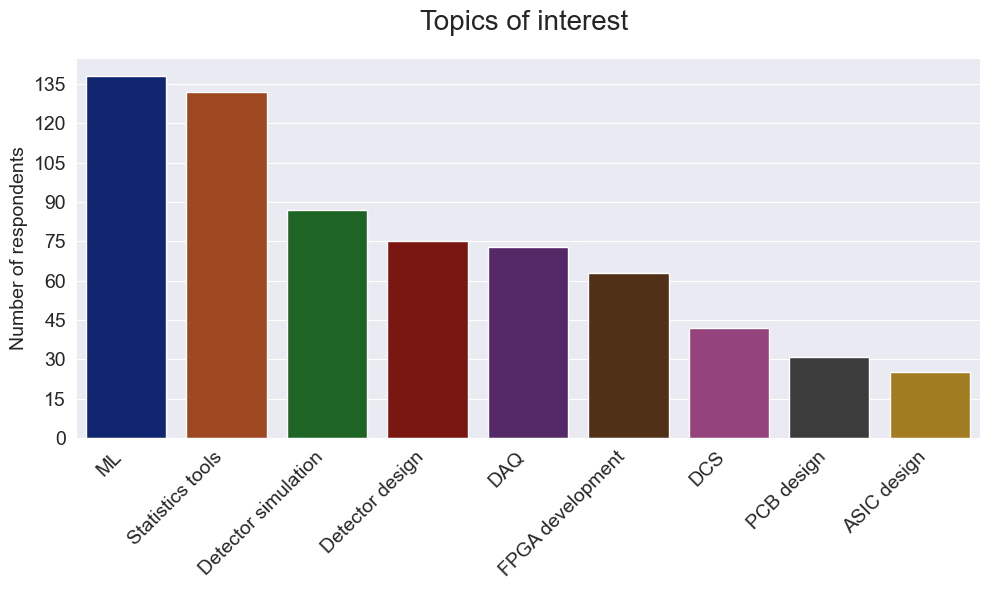}%
  }%
}
\setlength{\twosubht}{\ht\twosubbox}


\centering

\subcaptionbox{\label{fig:in:field}}{%
  \includegraphics[width=0.7\linewidth]{figures/introduction/barplot_What_is_your_field_of_research_You_may_select_more_than_one_option..png}%
}\quad
\subcaptionbox{\label{fig:in:topic}}{%
  \includegraphics[width=0.7\linewidth]{figures/introduction/barplot_Which_topics_are_you_interested_in_learning_about_You_may_select_more_than.png}%
}
\caption{The respondents’ fields of research distribution (a). Hot topics indicated by respondents (b).}
\end{figure}

The respondents’ fields of research are shown in Figure~\ref{fig:in:field}. The majority indicated working in data analysis. Other frequently mentioned areas included software and detector design and development. This distribution reflects the strong computational and analytical focus of the surveyed research community. In the last part of the general questions, we aimed to identify which skills are currently of greatest interest to ECRs; the results can be found in the Figure~\ref{fig:in:topic}. Taking into account the growing popularity of machine learning techniques and the multitude of possibilities of their application, as well as the fact that the respondents mainly work in data analysis, the most popular topics to learn are machine learning and advanced statistical techniques used in data analysis.

%% file: schools_general.tex
\begin{figure}[h!]
\sbox\twosubbox{%
  \resizebox{\dimexpr.99\textwidth-1em}{!}{%
    \includegraphics[height=3cm]{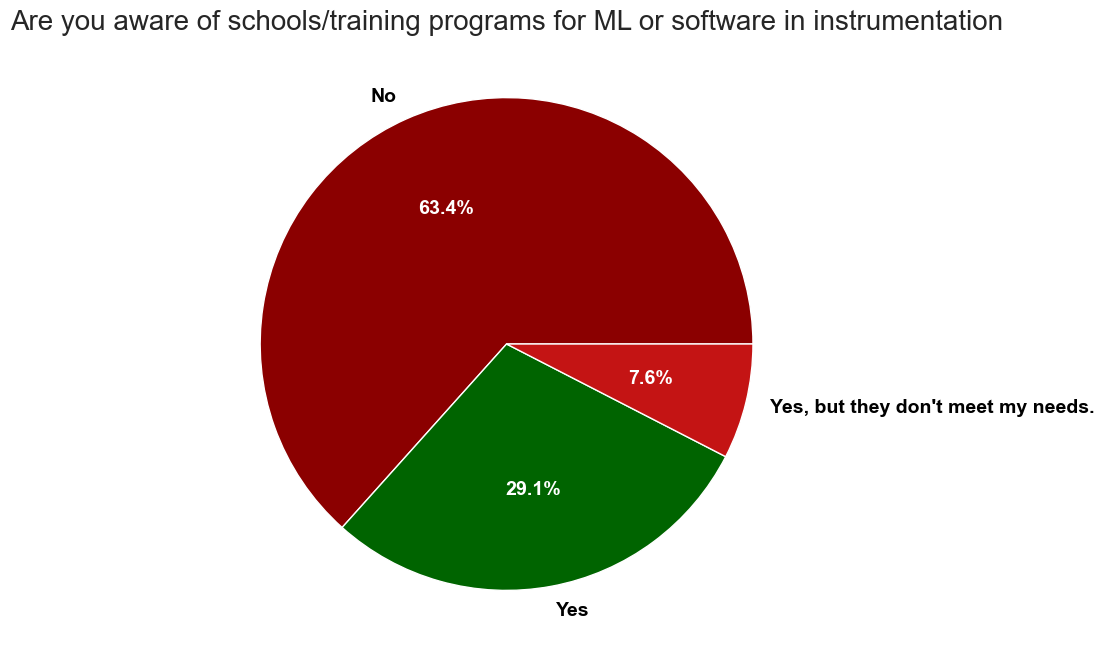}%
    \includegraphics[height=3cm]{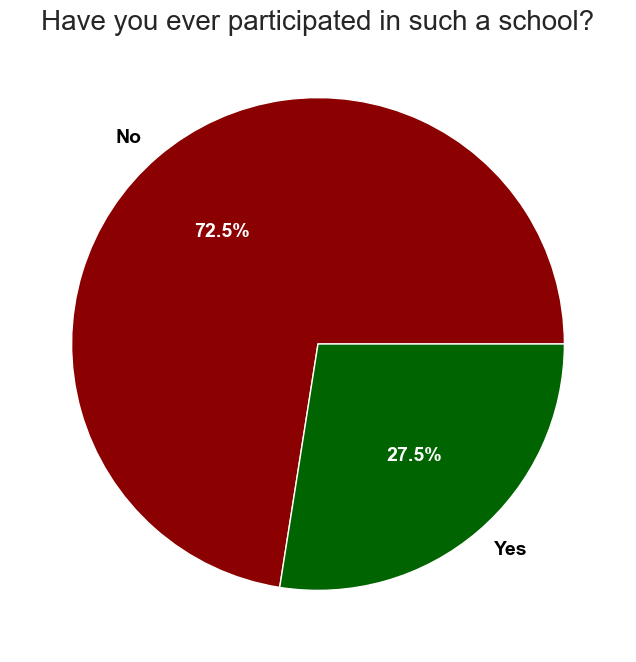}
  }%
}
\setlength{\twosubht}{\ht\twosubbox}


\centering

\subcaptionbox{\label{fig:in:knownOfShools}}{%
  \includegraphics[height=\twosubht]{figures/schoolsGeneral/Do_you_know_of_any_schools_or_training_programs_for_machine_learning_or_s.png}%
}\quad
\subcaptionbox{\label{fig:in:participatedInSchool}}{%
  \includegraphics[height=\twosubht]{figures/schoolsGeneral/Have_you_ever_participated_in_such_a_school.png}%
}
\caption{Knowledge of available training programs (a). Have participants attended such programs? (b)}
\end{figure}

In the first part of the survey, we examined participants’ awareness of the available training courses and their accessibility, as well as the quality of the existing schools. We also sought to identify gaps in the programs and other obstacles that might hinder participation in the schools or workshops. In Figure~\ref{fig:in:knownOfShools} we can see that access to information about training and schools remains a significant challenge. More than half of the respondents are unaware of the available training opportunities. Additionally, 7$\%$ of respondents indicated that, although they are aware of school offerings, the programs do not meet their expectations. The lack of knowledge may stem from the fact that there are no accessible websites that allow users to easily browse schools and filter them by category or required level of knowledge.
\\

\begin{figure}[h!]
\sbox\twosubbox{%
  \resizebox{\dimexpr.99\textwidth-1em}{!}{%
    \includegraphics[height=3cm]{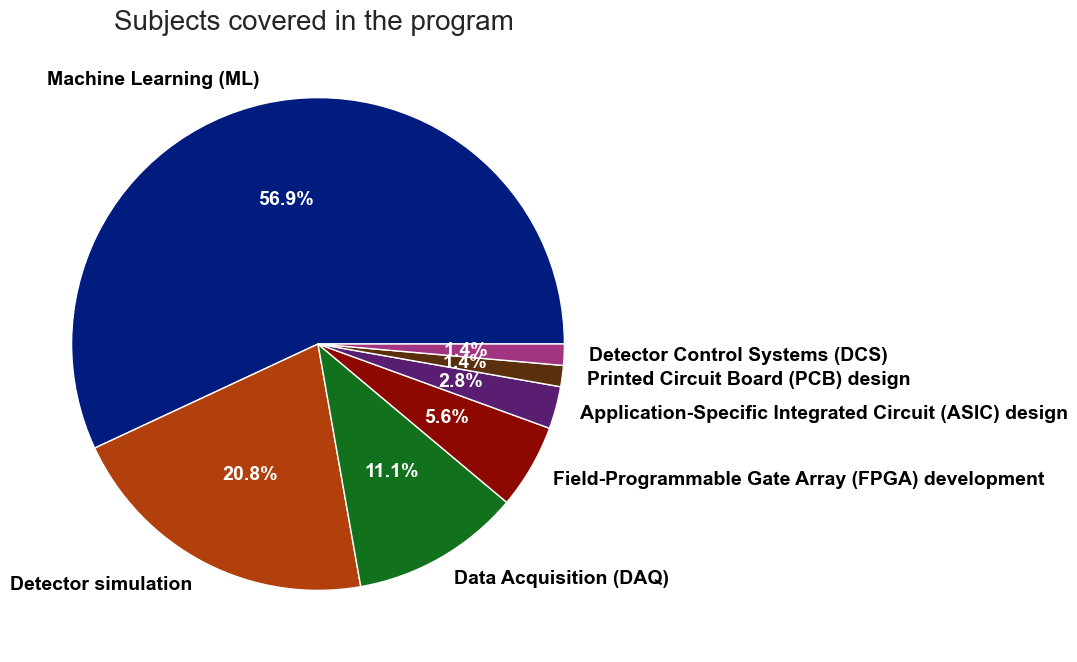}%
    \includegraphics[height=3cm]{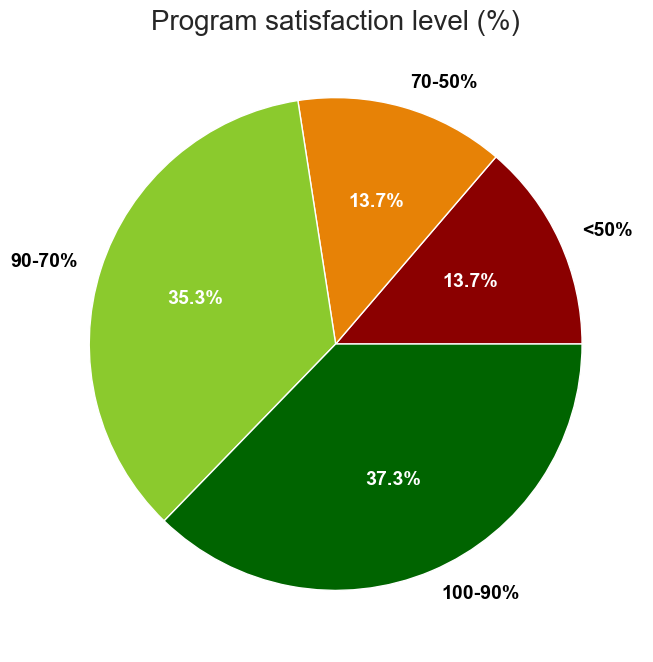}%
  }%
}
\setlength{\twosubht}{\ht\twosubbox}


\centering

\subcaptionbox{\label{fig:in:topiSchool}}{%
  \includegraphics[height=\twosubht]{figures/schoolsGeneral/If_yes,_what_topics_did_they_cover_You_may_select_more_than_one_option..png}%
}\quad
\subcaptionbox{\label{fig:in:satisfactionSchool}}{%
  \includegraphics[height=\twosubht]{figures/schoolsGeneral/Were_you_satisfied_with_the_program_you_attended_If_so,_please_specify_yo.png}%
}
\caption{Coverage of topics in attended programs (a). Participant satisfaction with training programs (percentage) (b).}
\end{figure}

Only 28.7$\%$ of respondents have participated in such a school, as shown in Figure~\ref{fig:in:participatedInSchool}. Among those who attended a school, more than half participated in a machine learning school (56.9$\%$). The second largest group attended Detector Simulation workshops (20.8$\%$), followed by participants in Data Acquisition (DAQ) workshops (11.1$\%$). Satisfaction with the schools is shown in Figure~\ref{fig:in:satisfactionSchool}, with over 72$\%$ of respondents reporting a satisfaction level above 70$\%$. 
\\

We also provided respondents with an open-ended question, giving them the opportunity to describe what they felt was missing in the school they attended, as well as what they considered to be the main challenges in participating in such training. Although we received only 21 responses out of 174 participants, several recurring themes emerged. Respondents noted that some courses could feel too advanced or fast-paced, with material often assuming prior knowledge that beginners did not always have. A need for more practical, hands-on experience was also frequently mentioned. Several participants felt that the schools were sometimes too theoretical, offering limited examples directly relevant to HEP analyses or providing short and demanding hands-on sessions.
\\

Participants additionally expressed interest in more structured learning paths that gradually progress from basic to advanced topics, supported by shorter but more frequent lessons, homework assignments, and more HEP-focused examples. Finally, several participants pointed to time-zone problems and restricted access to computing resources for practice after the course as factors that further limited their ability to fully benefit from the training. While the programs were generally appreciated for the expertise of the instructors and the usefulness of the content, respondents indicated that slower pacing, more complete tutorials, and content tailored to the audience would further enhance the learning experience.

%% file: 02_ML.tex
Machine learning (ML) techniques have become increasingly important in research across numerous scientific domains, even more so in High Energy Physics (HEP). As the complexity and volume of data continue to grow, the HEP community is turning to ML to enhance data analysis, simulation, and interpretation. However, the pathways through which these scientists acquire ML skills—and their perspectives on the most effective modes of learning—remain underexplored. Understanding how young researchers currently engage with ML, including the resources they use and the challenges they face, is crucial for tailoring educational programs that meet their evolving needs. The insights gained are intended to inform the organizers of future schools and workshops, helping to design targeted, effective training programs that better support the next generation of HEP scientists in mastering machine learning. This section is organized as follows: first, we review the current use cases of machine learning within the community, along with the software tools commonly employed. The second subsection explores the preferences and learning approaches of young researchers regarding ML education. Finally, we discuss how young researchers would organize a well-balanced curriculum for future machine learning schools.\\

\begin{figure}[h!]
\sbox\twosubbox{%
  \resizebox{\dimexpr.99\textwidth-1em}{!}{%
    \includegraphics[height=3cm]{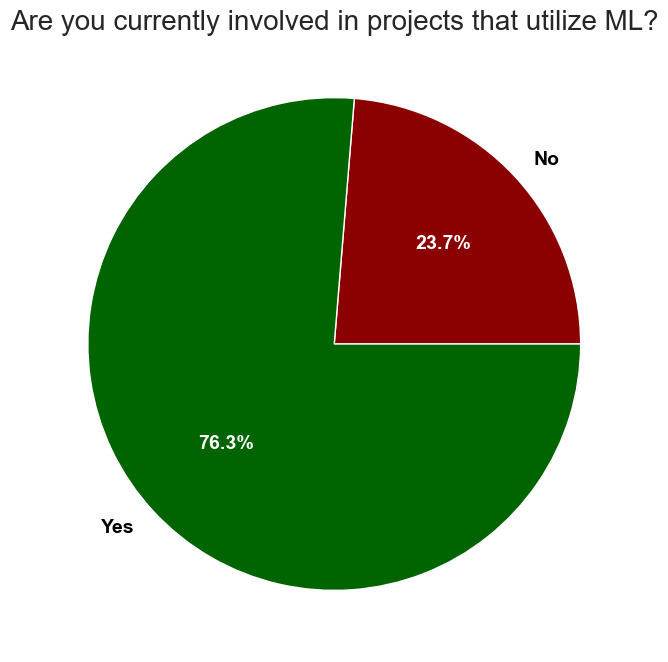}%
    \includegraphics[height=3cm]{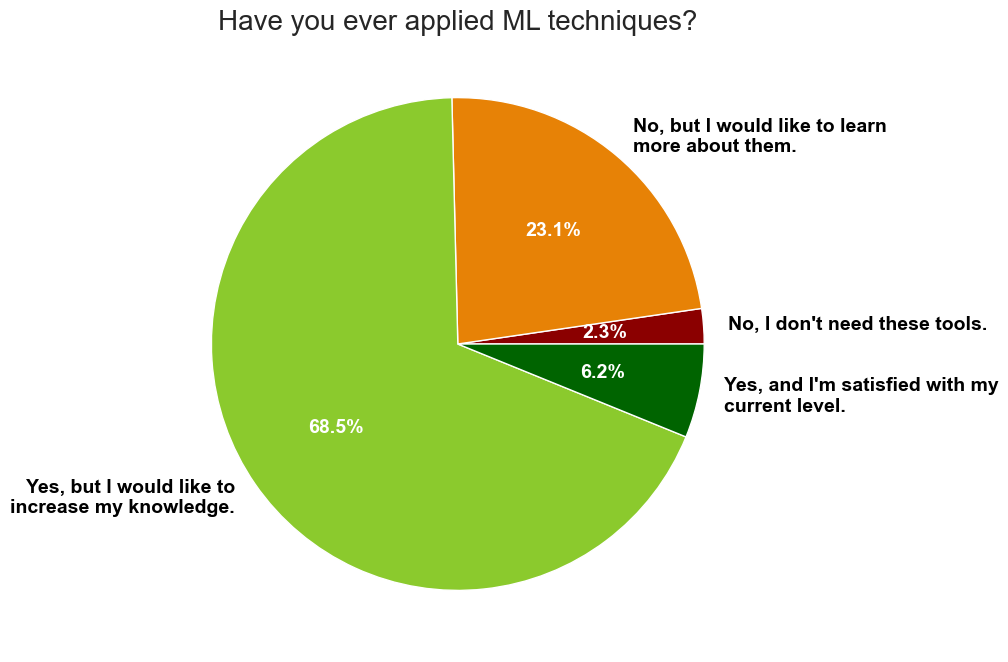}%
  }%
}
\setlength{\twosubht}{\ht\twosubbox}


\centering

\subcaptionbox{\label{fig:ML:useMLa}}{%
  \includegraphics[height=\twosubht]{figures/ML/Are_you_currently_involved_in_projects_that_utilize_ML,_or_are_you_interested_in_expl.png}%
}\quad
\subcaptionbox{\label{fig:ML:useMLb}}{%
  \includegraphics[height=\twosubht]{figures/ML/Have_you_ever_applied_ML_techniques.png}%
}
\caption{Usage of ML software (a) and knowledge level (b).}
\label{fig:ML:useML}
\end{figure}
\begin{figure}[h!]
\sbox\twosubbox{%
  \resizebox{\dimexpr.99\textwidth-1em}{!}{%
    \includegraphics[height=3cm]{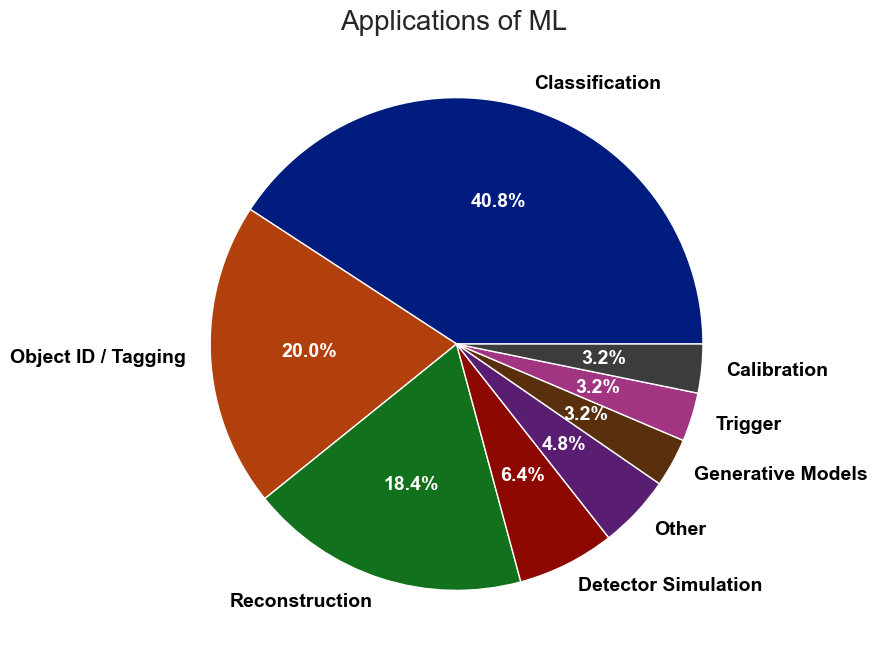}%
    \includegraphics[height=3cm]{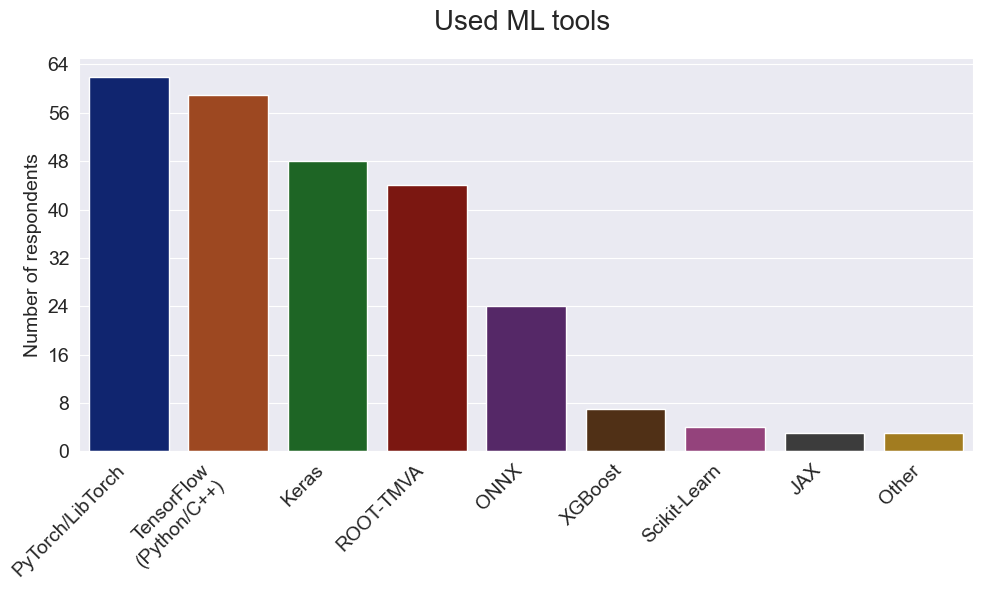}%
}
}
\setlength{\twosubht}{\ht\twosubbox}


\centering

\subcaptionbox{\label{fig:ML:whatMLArcha}}{%
  \includegraphics[height=\twosubht]{figures/ML/If_yes,_what_are_they_used_for_Please,_write_a_short_list,_separating_the_items_with.png}%
}\quad
\subcaptionbox{\label{fig:ML:whatMLArchb}}{%
  \includegraphics[height=\twosubht]{figures/ML/barplot_Which_ML_tools_do_you_use_in_your_work_You_may_select_more_than_one_option..png}%
}
\caption{The main use cases for ML (a). The ML libraries that are most often used  by respondents (b). }
\end{figure}

As can be seen in  Figure~\ref{fig:ML:useML}, a large majority of survey respondents are actively using ML or show interest in the field. Only 2\% of people claim to not need any ML tools in their work, highlighting the importance of ML in the field of HEP. More than 90$\%$ of respondents would however like to learn more of the subject, while only 6$\%$ are satisfied with their current knowledge of the subject. Figure~\ref{fig:ML:whatMLArcha} highlights the main use cases for ML in our field. We see the vast majority of use cases are focused on Classification tasks, both on the event level as well as on the object level (e.g. jet tagging). ML is often also applied on reconstruction tasks such as clustering or tracking but also including other regression tasks. Few respondents use machine learning for other goals. The survey also asked what ML tools we employ
in our work. Figure~\ref{fig:ML:whatMLArchb} shows that mainly Python-based ML libraries are used. While most respondents use PyTorch, the survey shows that the two other well-known frameworks Keras and its base library, TensorFlow, are also often employed. Training for all three libraries will hence remain of use to the community in the near future. We see that ROOT-TMVA, the machine learning library in the ROOT software framework developed at CERN is also used. ML packages that do not include neural network functionalities are less employed. Scikit-Learn and XGBoost are the libraries that are most often used for those cases.\\

\begin{figure}[h!]
\centering
  \includegraphics[width=0.63\linewidth]{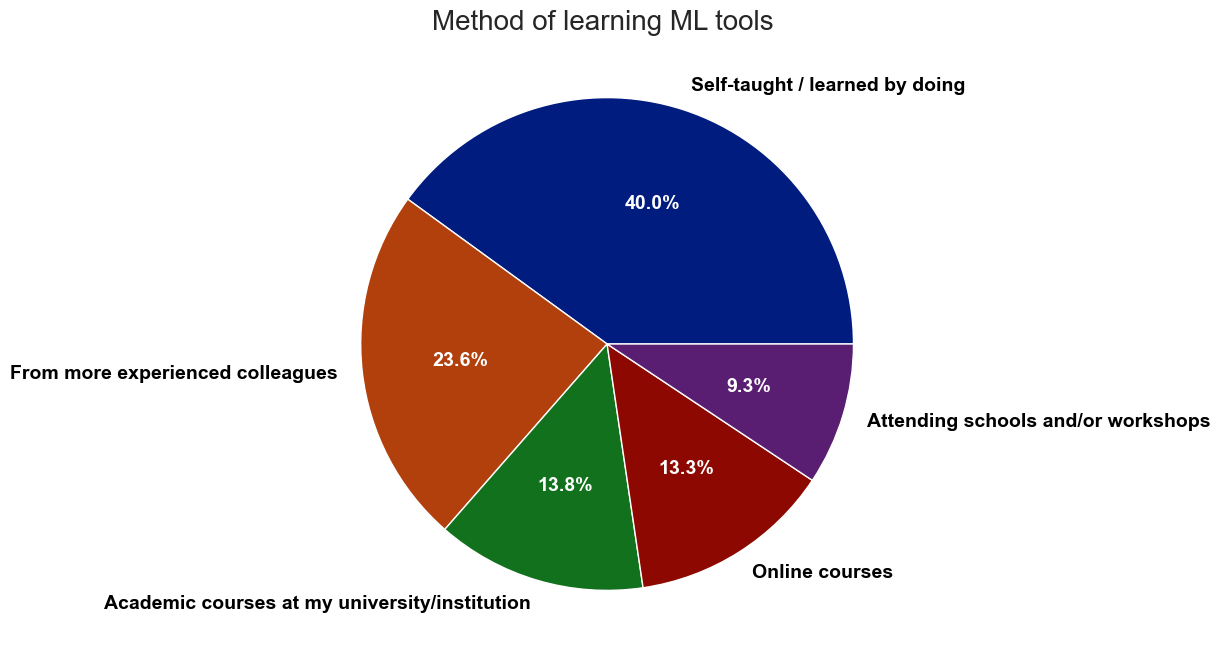}
\caption{Method of learning ML.}
\label{fig:ML:wherelearn}
\end{figure}

Roughly two thirds of respondents learned their ML techniques either independently or by getting help from more experienced colleagues. We see that roughly 13\% take courses at their university, while another 13\% refined their ML knowledge via online courses. The remaining 9\% of respondents have learned via participation to a ML school. While the number of people learning via participation to dedicated schools is quite low, it is higher than for any of the other tools investigated in the survey. These numbers are highlighted in Figure~\ref{fig:ML:wherelearn}.

\begin{figure}[h!]
\sbox\twosubbox{%
  \resizebox{\dimexpr.99\textwidth-1em}{!}{%
    \includegraphics[height=3cm]{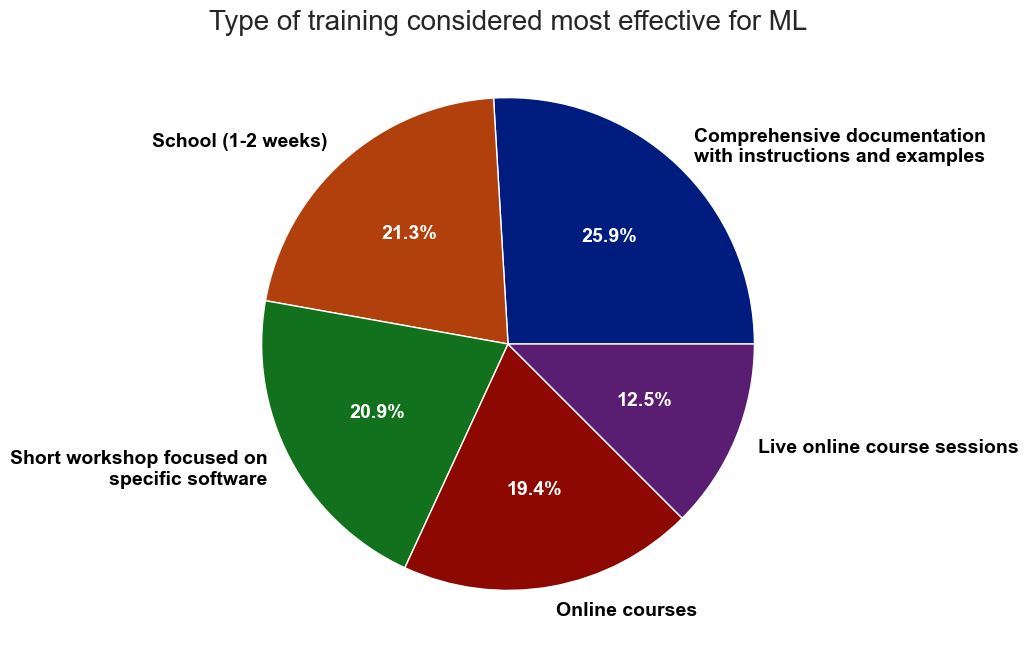}%
    \includegraphics[height=3cm]{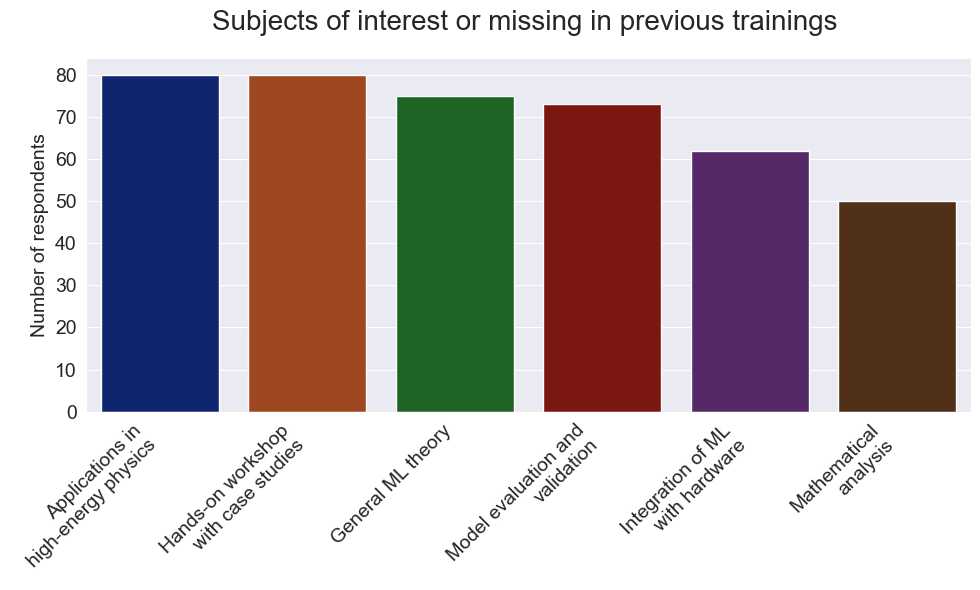}%
}
}
\setlength{\twosubht}{\ht\twosubbox}


\centering

\subcaptionbox{\label{fig:ML:MLschoola}}{%
  \includegraphics[height=\twosubht]{figures/ML/Which_kind_of_training_do_you_think_would_be_more_effective_for_ML.png}%
}\quad
\subcaptionbox{\label{fig:ML:MLschoolb}}{%
  \includegraphics[height=\twosubht]{figures/ML/barplot_What_would_you_like_to_learn,_or_what_do_you_feel_was_missing_from_the_training_you_r.png}%
}
\caption{The type of training considered most effective for ML (a). Subjects that are of interest to respondents or were underexplored in previous schools (b).}
\label{fig:ML:effective}
\end{figure}

When asked what type of training would be more effective for ML, the answers appear very evenly distributed. The most prominent answer, given by 25\% of respondents, is having a comprehensive documentation available with instructions and examples. Both organizing schools and organizing shorter workshops receive roughly 21\% of the votes each. The remaining 32\% of respondents would prefer to have online courses, where 40\% of those respondents want to have these courses live. When asked what aspects of the training they would want to receive or what part was missing, most responses were application and hands-on oriented, while the mathematical background and specific integration onto hardware scored lower. These numbers are shown in Figure~\ref{fig:ML:effective}.\\

\begin{figure}[h!]
    \centering
    
    \centering
    \includegraphics[width=0.53\linewidth]{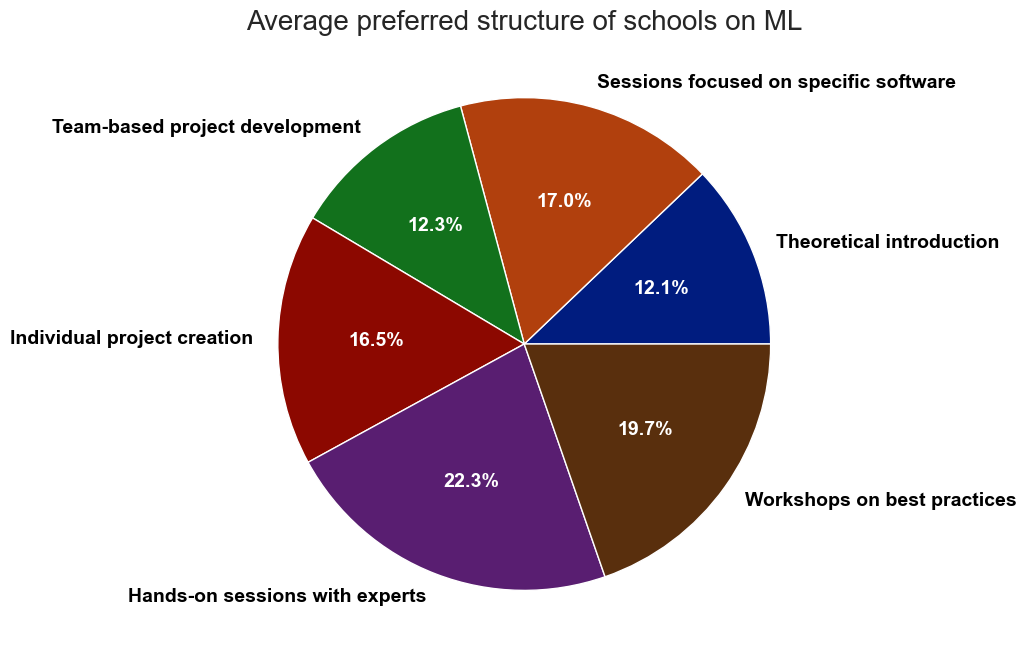}
    
\caption{The average preferred composition of a course on ML.}
\label{fig:ML:MLcompositionf}
\end{figure}

Survey participants were asked to compose their preferred structure for a ML school. They could select 6 topics and attribute a percentage of time to attribute to one subject. All responses were gathered and the average score per topic is shown in Figure~\ref{fig:ML:MLcompositionf}. Respondents attributed less time to the theoretical foundations of ML and time to work on team-based projects. Instead, hands-on sessions with experts and workshops on best practices received more priority.

%% file: 03_DetSim.tex

\begin{figure}[h!]
\sbox\twosubbox{%
  \resizebox{\dimexpr.99\textwidth-1em}{!}{%
    \includegraphics[height=3cm]{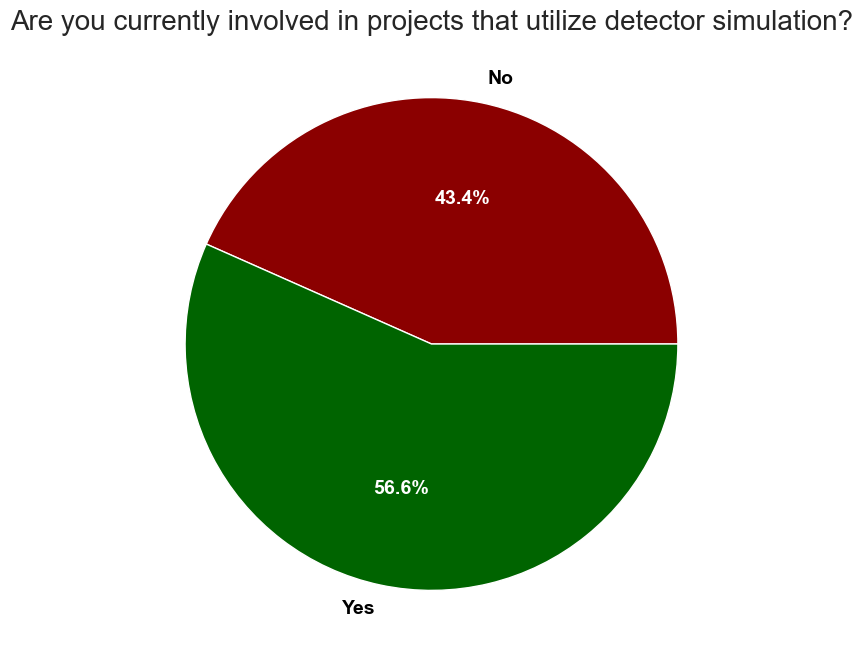}%
    \includegraphics[height=3cm]{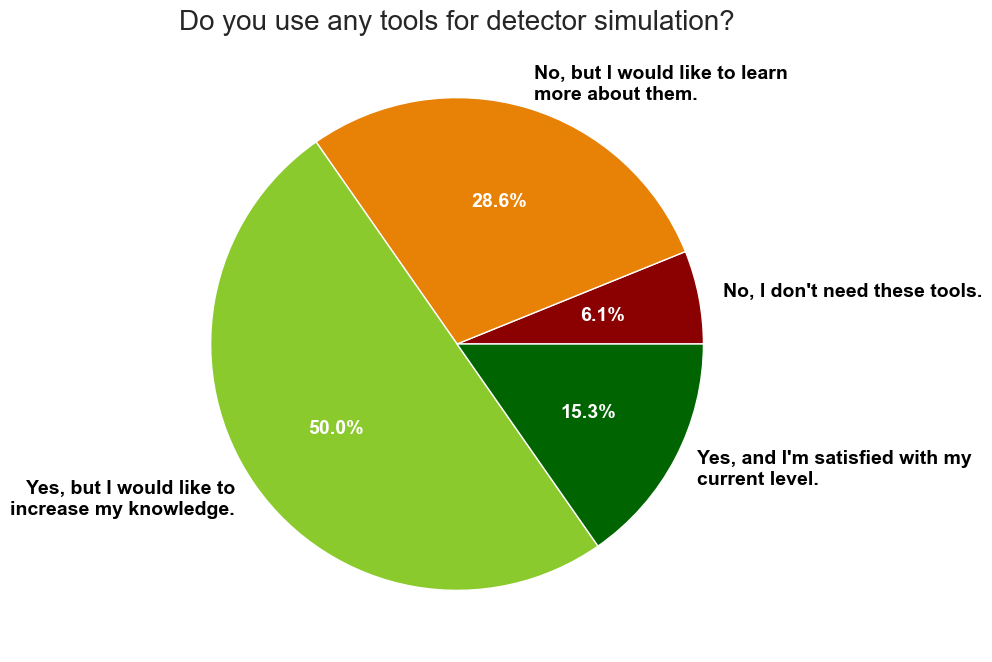}%
  }%
}
\setlength{\twosubht}{\ht\twosubbox}


\centering

\subcaptionbox{\label{fig:ML:usedetsima}}{%
  \includegraphics[height=\twosubht]{figures/ML/Are_you_currently_involved_in_projects_that_utilize_detector_simulation,_or_are_you_i.png}%
}\quad
\subcaptionbox{\label{fig:ML:usedetsimb}}{%
  \includegraphics[height=\twosubht]{figures/ML/Do_you_use_any_tools_for_detector_simulation.png}%
}
\caption{Usage of detector simulation software tools (a) and knowledge level (b).}
\end{figure}

Moving to the second HEP tools discussed in the survey, the experience of the respondents in using detector simulation tools was investigated. Respondents with detector simulation experience was notably lower at 56\%, as shown in Figure~\ref{fig:ML:usedetsima}. Notable is however that people with prior experience in detector simulation have a higher probability of being satisfied with their current level of knowledge on the tools, as shown in Figure~\ref{fig:ML:usedetsimb}. Roughly 6\% of respondents report that they do not need these tools.\\

\begin{figure}[h!]
\sbox\twosubbox{%
  \resizebox{\dimexpr.99\textwidth-1em}{!}{%
    \includegraphics[height=3cm]{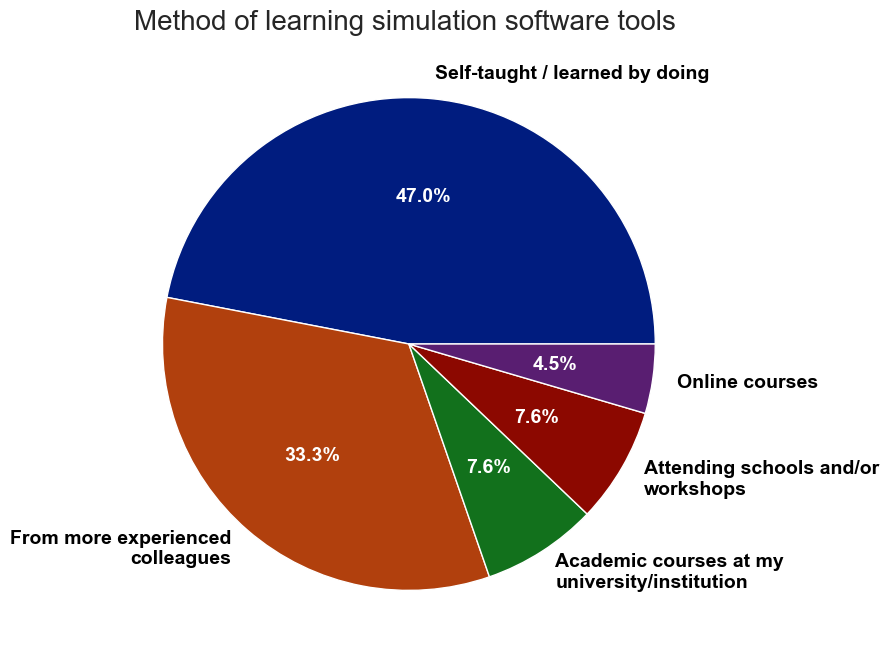}%
    \includegraphics[height=3cm]{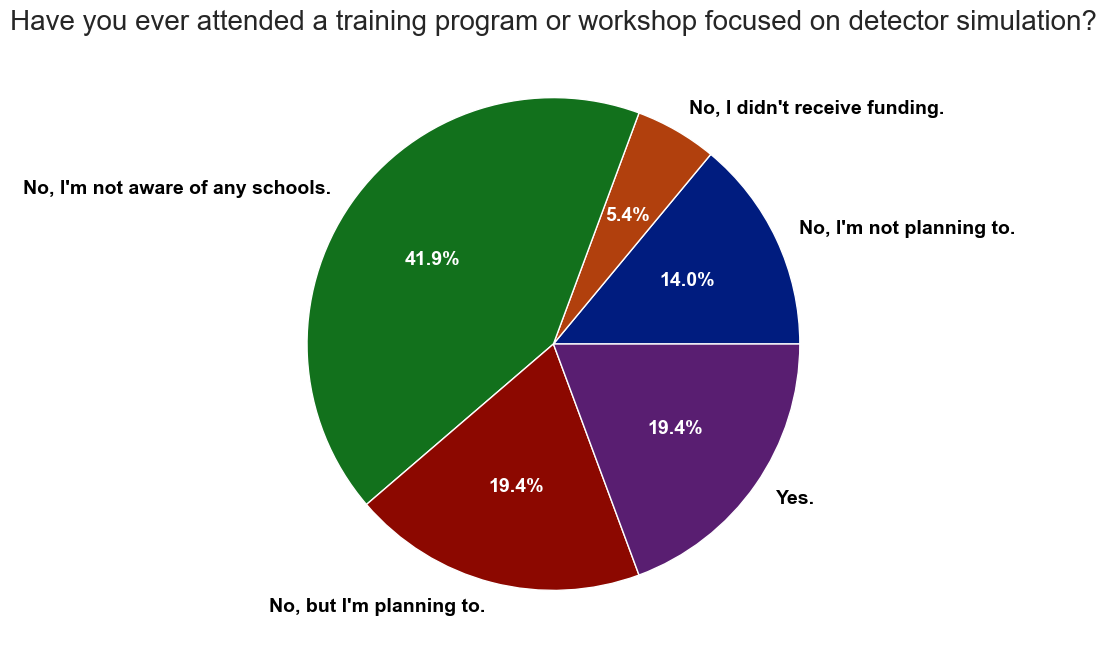}%
}
}
\setlength{\twosubht}{\ht\twosubbox}


\centering

\subcaptionbox{\label{fig:ML:wherelearndetsima}}{%
  \includegraphics[height=\twosubht]{figures/ML/If_you_are_currently_using_or_developing_simulation_software,_where_did_you_learn_to_.png}%
}\quad
\subcaptionbox{\label{fig:ML:wherelearndetsimb}}{%
  \includegraphics[height=\twosubht]{figures/ML/Have_you_ever_attended_a_training_program_or_workshop_focused_on_detector_simulation.png}%
}
\caption{Method of learning detector simulation software tools (a). Previous participation to a training program for detector simulation.}
\end{figure}

When asked how people acquired their knowledge on detector simulation tools, we see that roughly 80\%, notably more than is the case for ML, is trained by themselves or by experienced coworkers. Less people take courses from their institute, from organised schools or online, as shown in Figure~\ref{fig:ML:wherelearndetsima}. Figure~\ref{fig:ML:wherelearndetsimb} shows that when asked if people have attended a training program or workshop focused on detector simulation before, a large majority have not done so, while 20\% has. Another 20\% is still planning to, while a large fraction of 42\% is not aware of any schools on these subjects.

\newpage

\begin{figure}[h!]
\sbox\twosubbox{%
  \resizebox{\dimexpr.99\textwidth-1em}{!}{%
    \includegraphics[height=3cm]{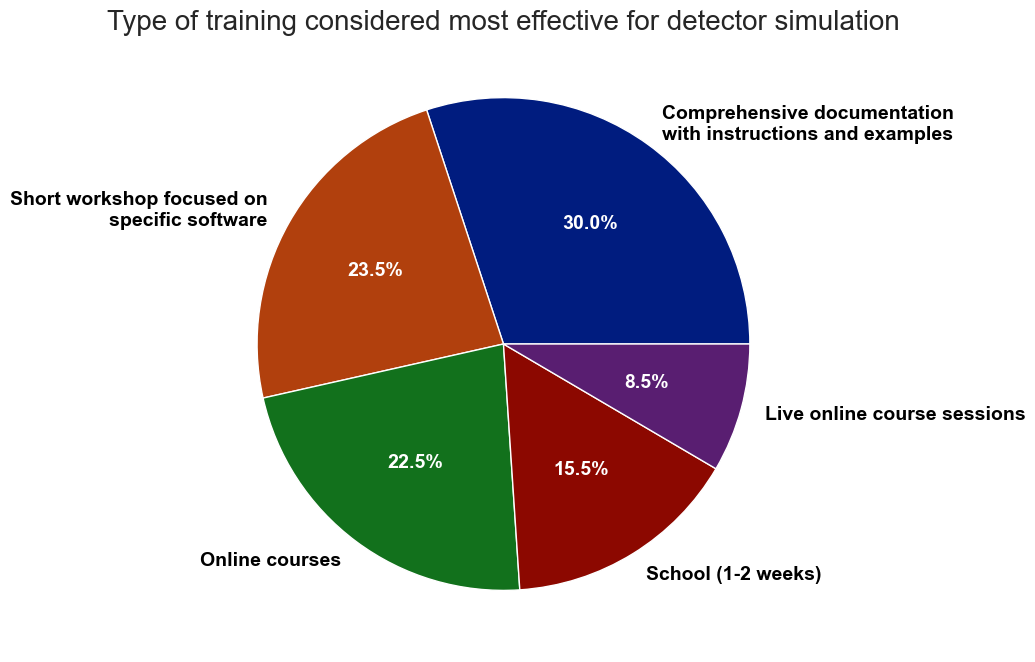}%
    \includegraphics[height=3cm]{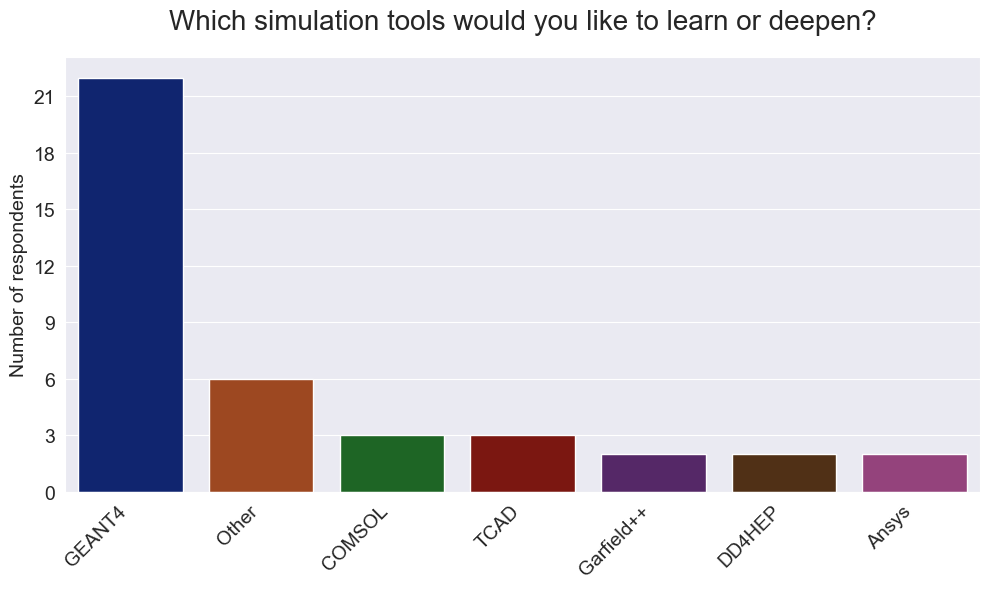}%
}
}
\setlength{\twosubht}{\ht\twosubbox}


\centering

\subcaptionbox{\label{fig:ML:detsimschoola}}{%
  \includegraphics[height=\twosubht]{figures/ML/What_type_of_training_do_you_think_would_be_most_effective_for_detector_simulation.png}%
}\quad
\subcaptionbox{\label{fig:ML:detsimschoolb}}{%
  \includegraphics[height=\twosubht]{figures/ML/barplot_Which_simulation_tools_would_you_like_to_learn_or_deepen_Please_list_them,_separatin.png}%
}
\caption{The type of training considered most effective for detector simulation (a). Simulation tools that are of interest to respondents
or were underexplored in previous schools (b).}
\end{figure}

When asked what type of training would be beneficial for learning to work with detector simulation tools, 30\% of respondents highlight comprehensive documentation with instructions and examples as the most effective tool. This is notably more than was the case for ML tools. Short workshops also get identified as an effective method 24\% of the time, while longer schools get selected as most effective 15\% of the time. This seems to indicate that teaching tools for detector simulation software are often preferred to be shorter compared to ML teaching methods. A similar amount (32\%) of respondents select online schools as an effective tool to teach detector simulation tools. Less respondents however see the need for these courses to be given live. These numbers are shown in Figure~\ref{fig:ML:detsimschoola}. When asked which tools respondents would like to learn more about, Geant4 clearly stands out as the most requested software, while other less requested detector simulation tools are also reported in Figure~\ref{fig:ML:detsimschoolb}.\\

\begin{figure}[h!]
    \centering
    \centering
    \includegraphics[width=0.56\linewidth]{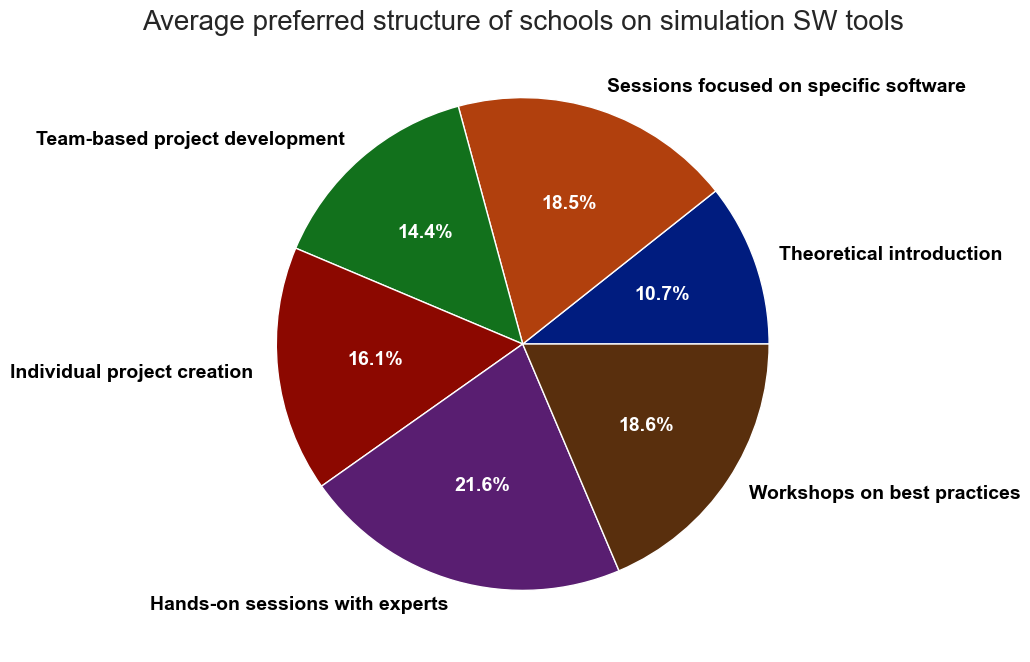}
\caption{The average preferred composition of a course on detector simulation.}
\label{fig:ML:detsimcomposition}
\end{figure}

Survey participants were asked to compose their preferred structure for a detector simulation software training school. They could select 6 topics and attribute a percentage of time to attribute to one subject. All responses were gathered and the average score per topic is shown in Figure~\ref{fig:ML:detsimcomposition}. Respondents attributed less time to the theoretical foundations of detector simulation and time to work on team-based projects. Instead, hands-on sessions with experts, workshops on best practices and sessions focused on specific software received more priority.\\

%% file: 04_DAQ.tex
High energy physics experiments produce enormous amounts of data when particles collide or interact with detectors.  
The Data Acquisition (DAQ) system is the chain of hardware and software components that selects, records, and prepares this information for later analysis. Detector Control Systems (DCS) are the other half of running a high-energy physics experiment, complementing the DAQ. While the DAQ handles physics data, the DCS handles detector safety, stability, and operation. For simplicity, both systems are going to be referred to as DAQ in the following.
\\

This section presents the survey results concerning how young researchers currently acquire knowledge about DAQ, their training needs, and their perspectives on the organization of DAQ courses or schools. The analysis is structured in the same way as for the previous subjects. First, we review the current use cases of DAQ within the community, together with the associated software tools. Second, we explore the preferences and learning approaches of young researchers regarding DAQ education. Finally, we consider how young researchers would organize a well-balanced curriculum for future detector DAQ schools.
\\

\begin{figure}[h!]
\sbox\twosubbox{%
  \resizebox{\dimexpr.99\textwidth-1em}{!}{%
    \includegraphics[height=3cm]{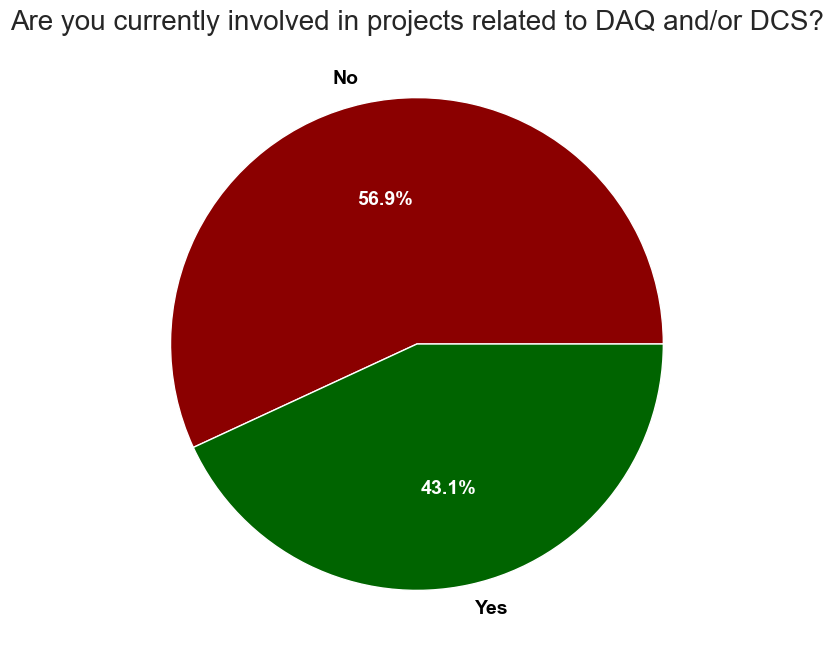}%
    \includegraphics[height=3cm]{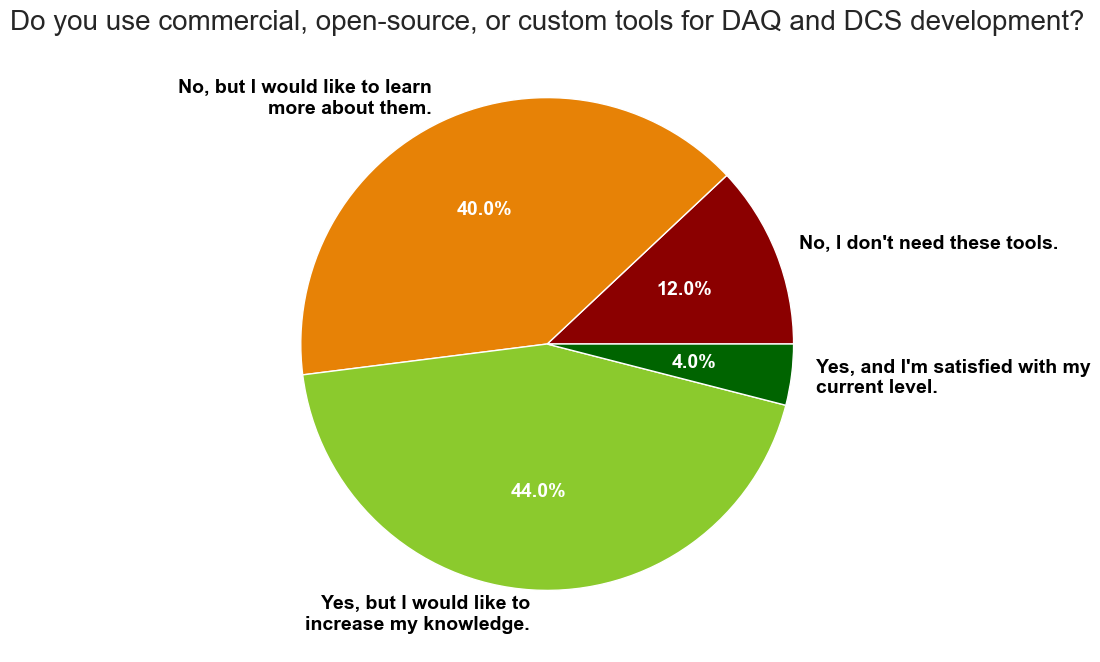}
}
}
\setlength{\twosubht}{\ht\twosubbox}


\centering

\subcaptionbox{\label{fig:DAQ:useDAQa}}{%
  \includegraphics[height=\twosubht]{figures/DAQ/Are_you_currently_involved_in_projects_related_to_DAQ_andor_DCS,_or_are_you_interest.png}%
}\quad
\subcaptionbox{\label{fig:DAQ:useDAQb}}{%
  \includegraphics[height=\twosubht]{figures/DAQ/Do_you_use_commercial,_open-source,_or_custom_tools_for_DAQ_and_DCS_development_e.g.png}%
}
\caption{
Involvement in DAQ related projects (a). Usage of DAQ software and knowledge level (b).}
\end{figure}

Figure~\ref{fig:DAQ:useDAQa} shows that 43\% of the survey respondents are currently involved in or interested in DAQ projects. The next questions of the survey are only answered by respondents that identified themselves as currently involved or interested in DAQ. Figure~\ref{fig:DAQ:useDAQb} shows the detailed results about the usage of commercial, open-source or custom tools for DAQ. The most common answer, with 44\%, is that ECRs use such software but they would like to increase their knowledge. Another 40\% of the respondents indicate their interest in learning about DAQ software, although they currently do not employ it. The fraction of respondents indicating that they do not need or employ such software is 12\%, and only 4\% of the respondents indicate their satisfaction with their current DAQ software knowledge. A follow-up question attempted to identify the most commonly used DAQ software tools, but received only a handful of responses.  \textsc{Grafana}, \textsc{Vivado}, and \textsc{LabView} are the tools more commonly quoted by the respondents.

\begin{figure}[h!]
\sbox\twosubbox{%
  \resizebox{\dimexpr.99\textwidth-1em}{!}{%
    \includegraphics[height=3cm]{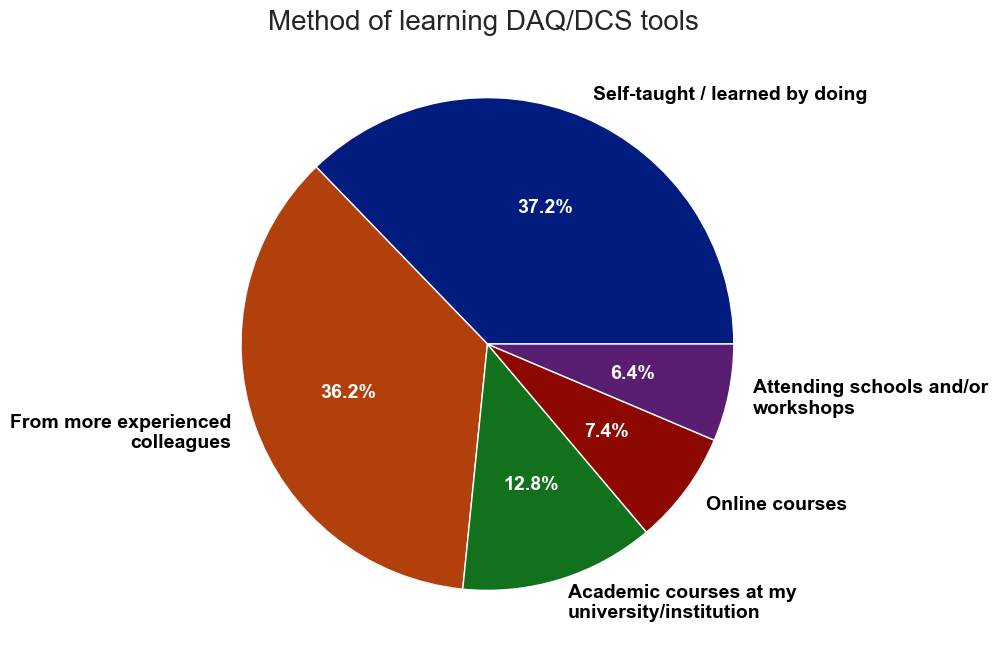}%
    \includegraphics[height=3cm]{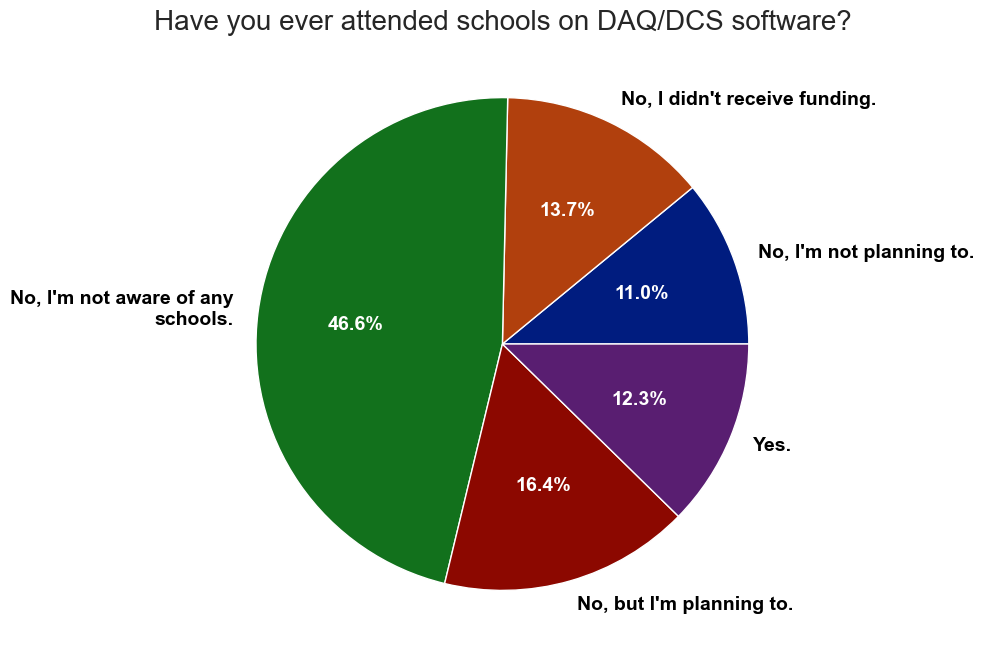}%
}
}
\setlength{\twosubht}{\ht\twosubbox}


\centering

\subcaptionbox{\label{fig:DAQ:formationDAQ}}{%
  \includegraphics[height=\twosubht]{figures/DAQ/If_you_are_currently_using_or_developing_DAQDCS_software,_where_did_you_learn_to_use.png}%
}\quad
\subcaptionbox{\label{fig:DAQ:schoolDAQ}}{%
  \includegraphics[height=\twosubht]{figures/DAQ/Have_you_ever_attended_schools_on_DAQDCS_software.png}%
}
\caption{Formation in DAQ (a) and DAQ school attendance (b).}
\end{figure}

The next questions of the survey targeted those respondents already using or interested in DAQ and DCS software. Figure~\ref{fig:DAQ:formationDAQ} shows how the respondents acquired their current knowledge: roughly three quarters of the respondents did it by themselves (37\%) or from more experienced colleagues (36\%). Only the  remaining quarter acquired it from some kind of organized course, either at their institutions (13\%), through online courses (7\%) or at a school/workshop (6\%). 
The second question of this block aims to identify if the respondents know the existence of DAQ software schools and their degree of participation. Figure~\ref{fig:DAQ:schoolDAQ} shows that half are not aware of any related school (47\%). Of the other half which is aware of DAQ software schools, half have either already attended a school (12\%), or are planning to attend one (16\%). The remaining quarter is split between respondents who cannot attend a school due to the lack of funding (14\%) and those who are not planning to attend any school (11\%).\\

\begin{figure}[h!]
\sbox\twosubbox{%
  \resizebox{\dimexpr.99\textwidth-1em}{!}{%
    \includegraphics[height=3cm]{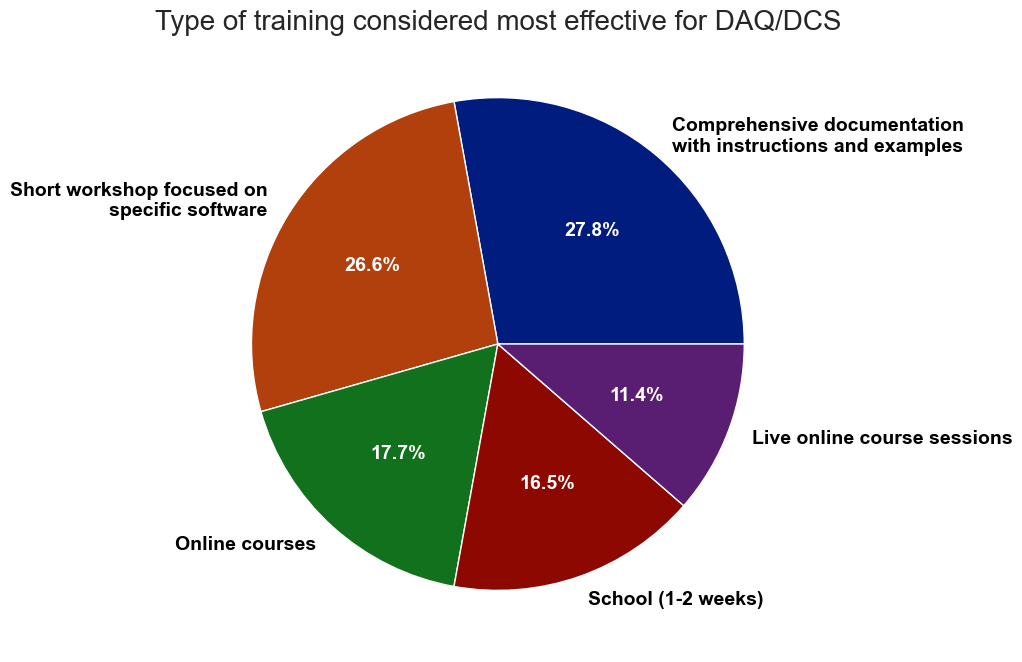}%
    \includegraphics[height=3cm]{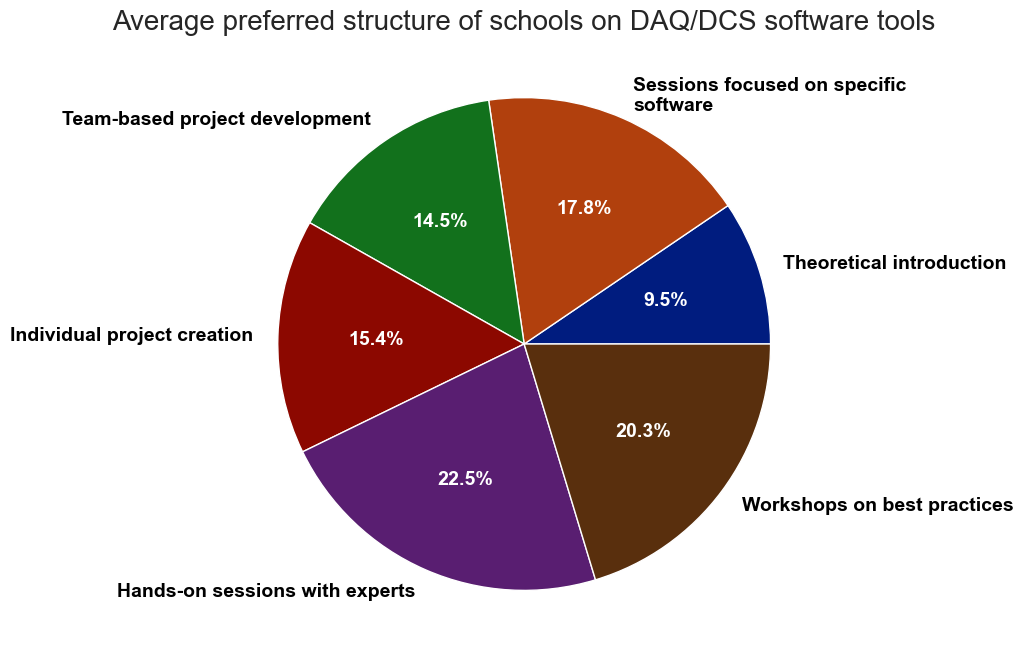}%
}
}
\setlength{\twosubht}{\ht\twosubbox}


\centering

\subcaptionbox{\label{fig:DAQ:formationprefDAQ}}{%
  \includegraphics[height=\twosubht]{figures/DAQ/What_type_of_training_do_you_believe_would_be_most_effective_for_DAQDCS.png}%
}\quad
\subcaptionbox{\label{fig:DAQ:schoolbalanceDAQ}}{%
  \includegraphics[height=\twosubht]{figures/DAQ/DAQstructure.png}%
}
\caption{DAQ software training preferences (a) and preferred school structure (b).}
\end{figure}

The last section of the survey targets the design of DAQ software training. When asked about their training preferences, the two most common answers (around 27\% each) are through comprehensive documentation and short focused workshops. Online courses and schools are seen each as the best option by around 17\% of the respondents and only 11\% prefer live online sessions. Detailed results are shown in Figure~\ref{fig:DAQ:formationprefDAQ}.
Figure~\ref{fig:DAQ:schoolbalanceDAQ} shows the relative weight that the respondents would assign to specific sections in a DAQ software school. There is a slight preference for hands-on sessions with experts and workshops on best practices, with around 20\% each. Sessions focused on specific software, individual and team-based development projects follow with roughly 15\% each. Finally, the theoretical introduction is the section with a smaller preferred weight (10\%). We find that these percentages can be used as guidelines when designing a school or course in DAQ software.

%% file: 05_Electronics.tex
Detector electronics are an essential ingredient of modern high-energy physics experiments. These experiments typically involve particle detectors, which measure properties such as the energy, momentum, position or time of flight of particles produced in high-energy collisions or in rare processes. Detector electronics bridge the gap between the physical detector and the digital domain, converting the weak analog signals generated by particle interactions into digital signals that can be processed, stored and analyzed.
Current HEP experiments employ detector electronics in multiple steps, including signal generation, amplification, digitization and triggering, noise filtering and monitoring and synchronization tasks. In addition, the continuous need for detector electronic improvements of HEP experiments drives continuous innovation in analog and digital electronics such as custom ASICs, high-speed data links, and advanced FPGA-based systems.
\\

This part of the survey provides insight into how young researchers engage with detector electronics and is structured in the same way as for the previous subjects. How do students currently acquire knowledge in this area, what training they consider most needed, and how they believe future courses or schools on detector electronics should be structured? To address these questions, we first outline the main use cases of detector electronics within the community and the software tools typically employed. We then examine the preferred learning approaches and educational expectations of young researchers in this field. The section concludes with a discussion of how respondents would design a balanced curriculum for future detector electronics schools.
\\

\begin{figure}[h!]
\sbox\twosubbox{%
  \resizebox{\dimexpr.99\textwidth-1em}{!}{%
    \includegraphics[height=3cm]{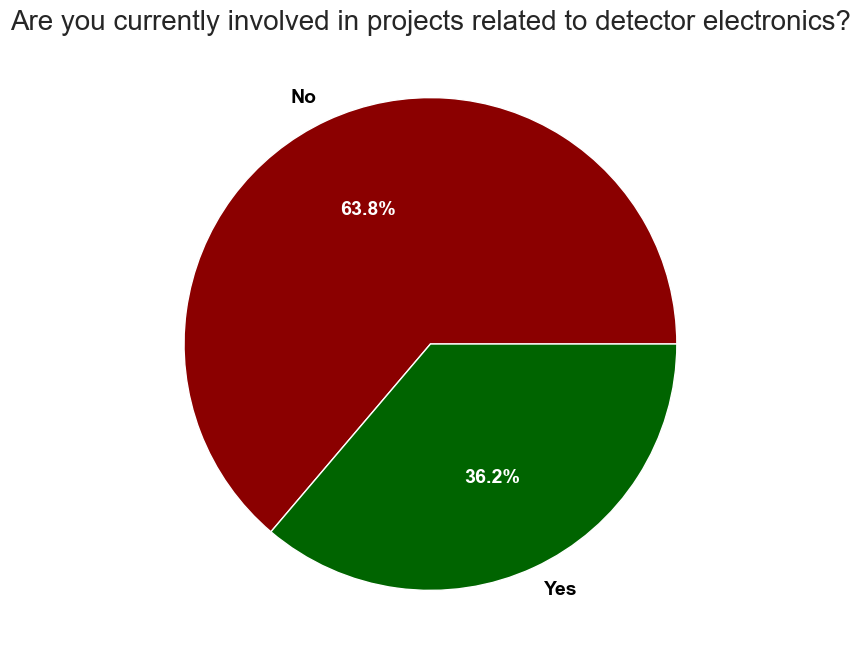}%
    \includegraphics[height=3cm]{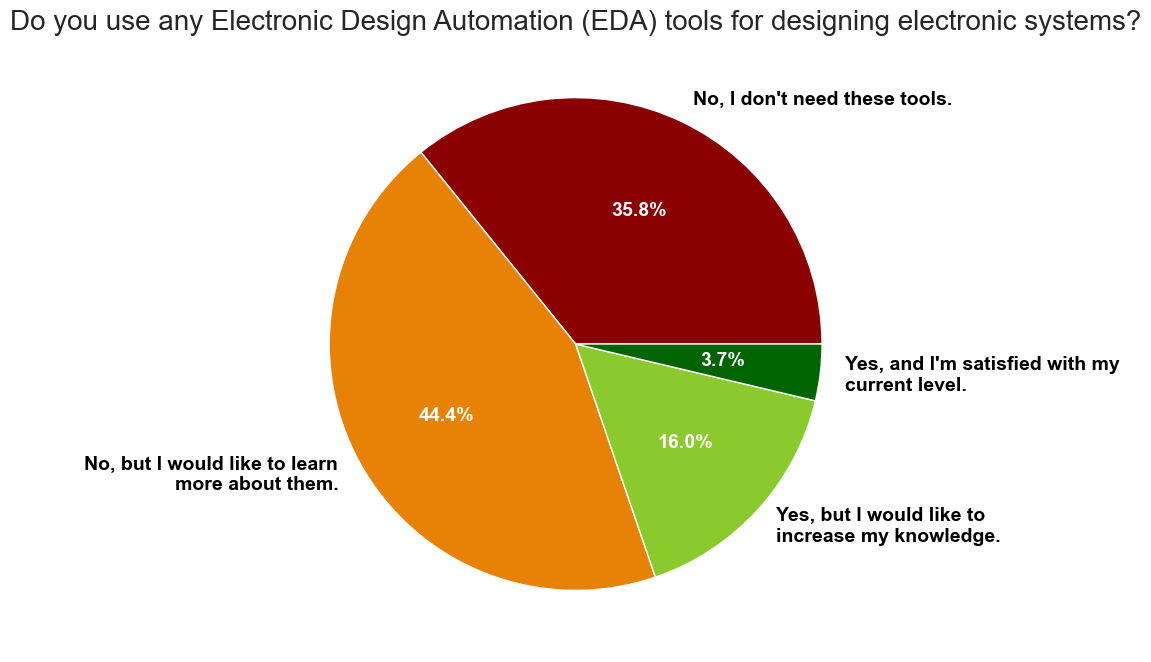}%
}
}
\setlength{\twosubht}{\ht\twosubbox}


\centering

\subcaptionbox{\label{fig:electronics:useEDAa}}{%
  \includegraphics[height=\twosubht]{figures/Electronics/Are_you_already_involved_in_projects_on_detector_electronics_or_are_you_inte.png}%
}\quad
\subcaptionbox{\label{fig:electronics:useEDAb}}{%
  \includegraphics[height=\twosubht]{figures/Electronics/Do_you_use_any_Electronic_Design_Automation_EDA_tools_for_designing_electr.png}%
}
\caption{Involvement in detector electronics related projects (a). Usage of detector electronics software and knowledge level (b). }
\end{figure}

\begin{figure}[h!]
    \centering
    \includegraphics[width=0.5\linewidth]{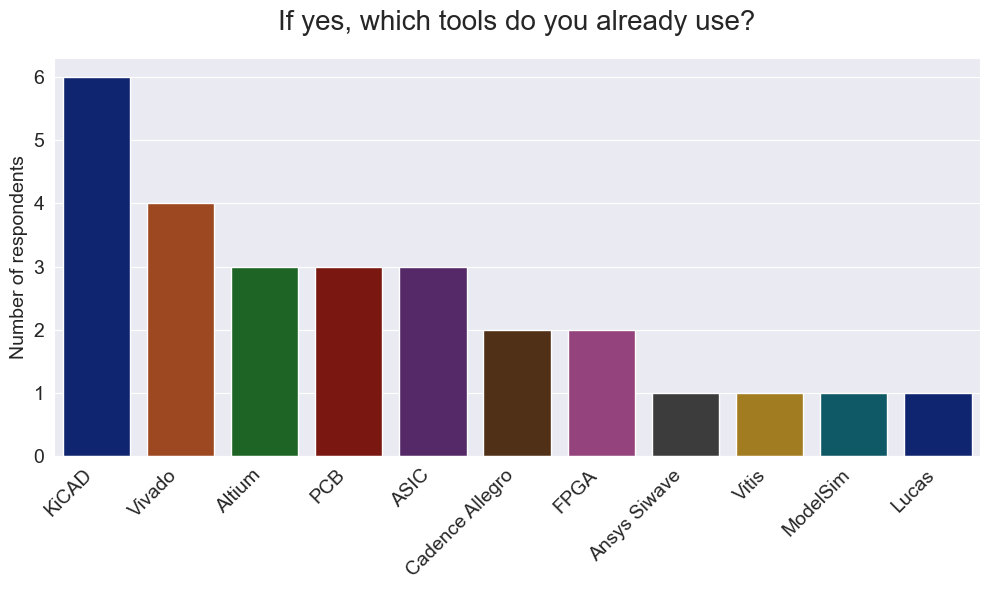}
    
\caption{Detector electronic software tools used by the survey respondents.}
\label{fig:electronics:commonEDAtools}
\end{figure}

Figure~\ref{fig:electronics:useEDAa} shows that only one third of the survey respondents are currently involved or interested in detector electronics projects. The following questions of the survey referring to electronics were answered only by a subset of the respondents. When asked about the usage of Electronic Design Automation (EDA) software, only half of the respondents provided an answer. The detailed results are shown in Figure~\ref{fig:electronics:useEDAb}. A third of the respondents indicated that they do not need or employ such software. Of the remaining respondents, two thirds indicate that they would like to learn about the related software despite they do not use it currently. Only around a fifth of the respondents currently employ EDA software, of which the vast majority would like to further increase their knowledge on the topic. A follow-up question attempted to identify the most commonly used EDA software tools, but received only limited responses. Results are shown in Figure~\ref{fig:electronics:commonEDAtools}.\\

\begin{figure}[h!]
\sbox\twosubbox{%
  \resizebox{\dimexpr.99\textwidth-1em}{!}{%
    \includegraphics[height=3cm]{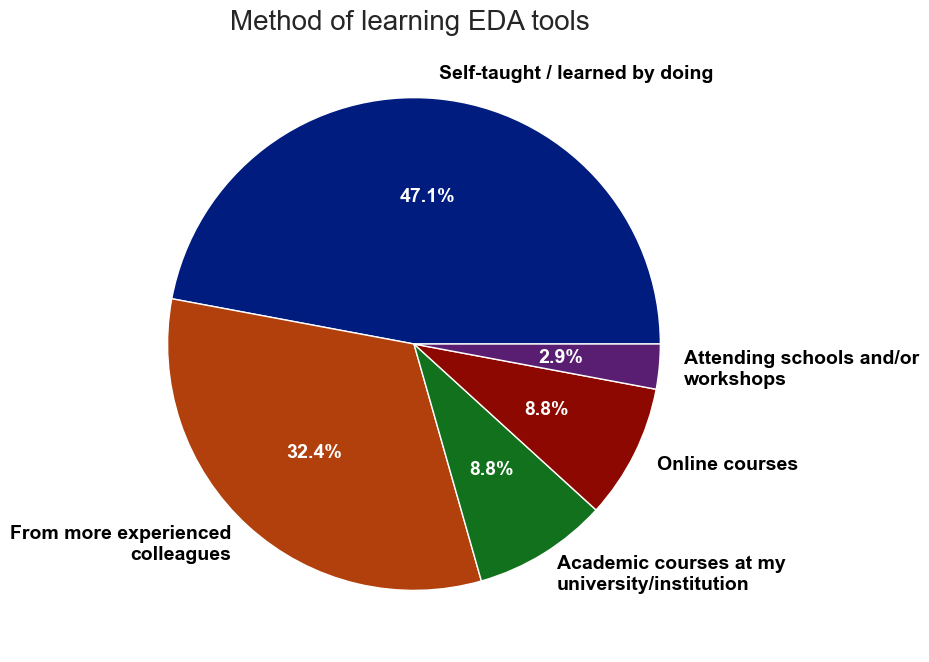}%
    \includegraphics[height=3cm]{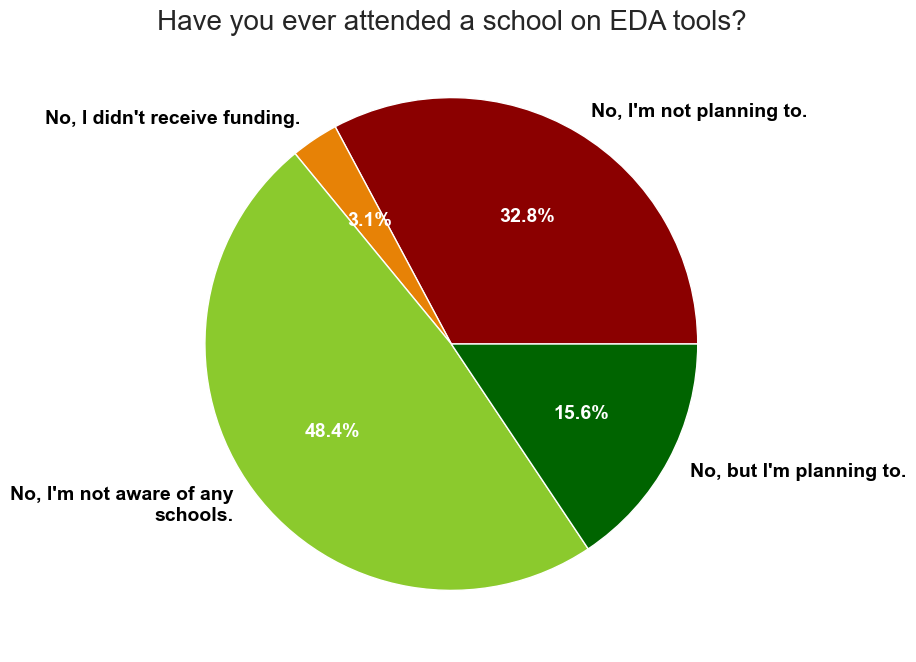}%
}
}
\setlength{\twosubht}{\ht\twosubbox}


\centering

\subcaptionbox{\label{fig:EDA:formationEDA}}{%
  \includegraphics[height=\twosubht]{figures/Electronics/If_you_are_currently_using_EDA_tools,_where_did_you_learn_to_use_them_You_m.png}%
}\quad
\subcaptionbox{\label{fig:EDA:schoolEDA}}{%
  \includegraphics[height=\twosubht]{figures/Electronics/Have_you_ever_attended_a_school_on_EDA_tools.png}%
}
\caption{Formation in detector electronics (a) and EDA software school attendance (b).}
\end{figure}

The next questions of the survey targeted those respondents already using or interested in EDA software. Figure~\ref{fig:EDA:formationEDA} shows how the respondents acquired their current knowledge: half of them by themselves, a third from more experienced colleagues, and only a fifth from dedicated courses or schools.
Figure~\ref{fig:EDA:schoolEDA} shows that half are not aware of any school related to EDA software. Of the other half, two thirds of the respondents declares that they are not planning to attend any of the available schools. Only 15\% of the respondents indicate that they are planning to attend a school on the topic. The fraction of respondents indicating lack of funding as the reason to not participate in a school is only 3\%. Here it is worth mentioning that Detector electronics is the only block in which none of the respondents has attended a related school. \\

\begin{figure}[h!]
\sbox\twosubbox{%
  \resizebox{\dimexpr.99\textwidth-1em}{!}{%
    \includegraphics[height=3cm]{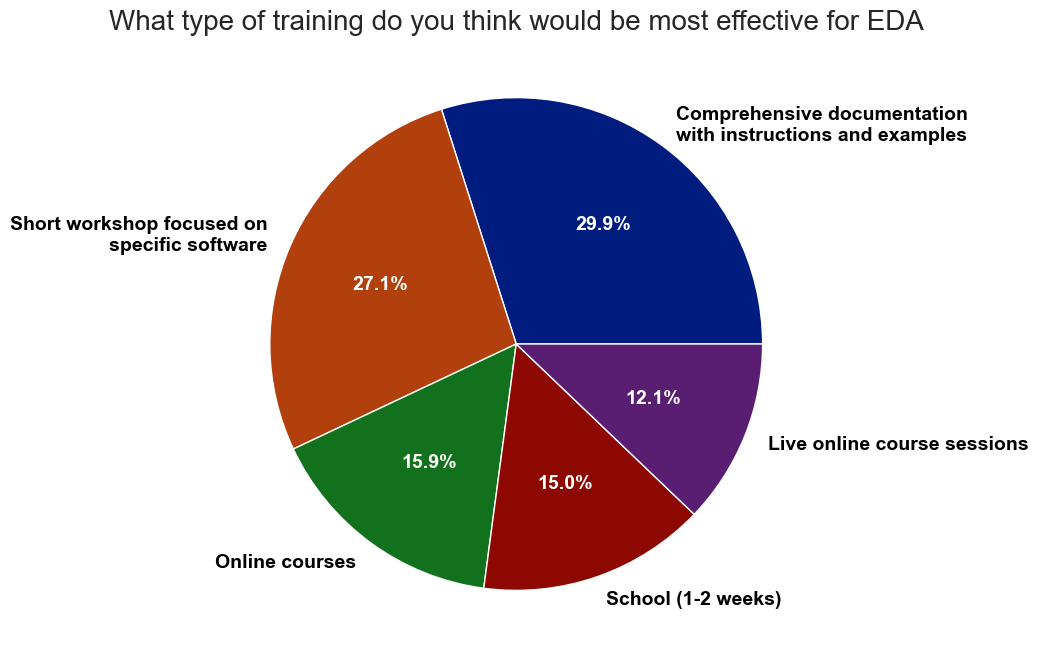}%
    \includegraphics[height=3cm]{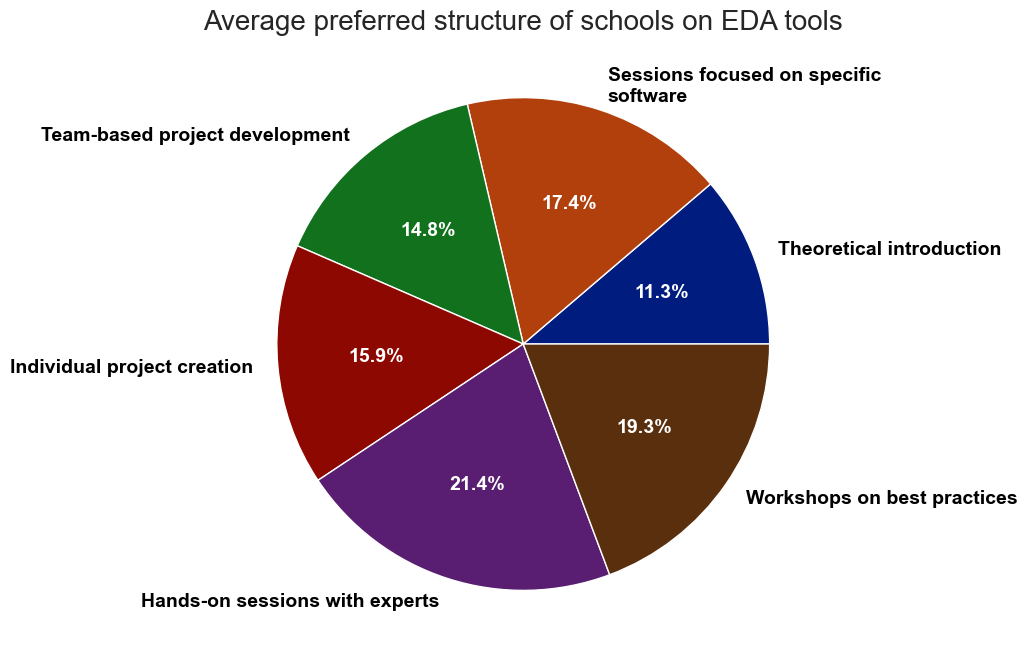}%
}
}
\setlength{\twosubht}{\ht\twosubbox}


\centering

\subcaptionbox{\label{fig:EDA:formationprefEDA}}{%
  \includegraphics[height=\twosubht]{figures/Electronics/What_type_of_training_do_you_think_would_be_most_effective_for_EDA.png}%
}\quad
\subcaptionbox{\label{fig:EDA:schoolbalanceEDA}}{%
  \includegraphics[height=\twosubht]{figures/Electronics/EDAstructure.png}%
}
\caption{Detector electronics training preferences (a) and preferred school structure (b).}
\end{figure}

When asked about their training preferences, the two most common answers (around 30\% each) are through comprehensive documentation and short focused workshops. Online courses and schools are seen each as the best option by around 15\% of the respondents and only 12\% prefer live online sessions. Detailed results are shown in Figure~\ref{fig:EDA:formationprefEDA}.
Figure~\ref{fig:EDA:schoolbalanceEDA} shows the relative weight that the respondents would assign to specific sections in a EDA software school. There is a slight preference for hands-on sessions with experts and workshops on best practices, with around 20\% each. Sessions focused on specific software, individual and team-based development projects follow with roughly 15\% each. Finally, the theoretical introduction is the section with a smaller preferred weight (10\%). We find that these percentages can be used as guidelines when designing a school or course in EDA software.

%% file: Conclusion.tex
This report summarizes the results of a survey carried out among european ECRs about training quality in instrumentation software and machine learning for HEP carried by the Software and Machine Learning for Instrumentation group of the ECFA ECRs panel. The survey targeted the training quality, preferences and needs in HEP, first with a general section about HEP software training, followed by four different blocks on specific topics: Machine learning; Detector simulation; Data Acquisition and detector control systems; and Detector electronics. A total of 174 responses were received, with varied participation in each of the blocks, Machine learning being the block receiving the largest number of responses and detector electronics the one with least participation.\\ 

The responses, which vary from block to block, are detailed in each of the corresponding sections, but some general patterns are observed and highlighted in this section. The first one is that in all blocks, the respondents that are eager to deepen their knowledge about the corresponding software toolsf of each category are the large majority. Another one is that the most common training method is self-taught, followed by learning from more experienced colleagues. Respondents taking part in dedicated schools, courses or workshops for HEP software training represent a minority in all blocks. Comprehensive documentation and short workshops on specific software are seen as the most effective type of training in all blocks. Finally, a similar pattern among blocks is observed among the respondents when asked about their preferred composition of a software school. There is a slight preference for hands-on sessions with experts and workshops on best practices, with around 20\% each. Sessions focused on specific software, individual and team-based development projects follow with roughly 15\% each. Finally, the theoretical introduction is the section with a smaller preferred weight (10\%). We find these percentages could be used as a reference when designing such training programs.\\

Given the generally low participation in dedicated schools and workshops, another recommendation from the authors is to make the training documentation of existing and future schools publicly available, so that it can be used by the community for offline and self-paced training. Respondents also point out that it is often difficult to find clear information about existing schools. Therefore, the authors suggest improving the visibility and accessibility of such initiatives, for example by creating a centralized website listing available courses with well-designed filters—including the level of each school—and by explicitly stating the expected level of prior knowledge in the school descriptions.